\documentclass[12pt,letterpaper]{article}
\usepackage{epsfig}
\usepackage{amsmath}
\usepackage{amssymb}
\usepackage{amsthm}
\usepackage{indentfirst}
\usepackage{xspace}
\usepackage{multirow}
\usepackage{hyperref}
\usepackage{xcolor}
\definecolor{darkred}{rgb}{0.5,0.15,0.15}
\hypersetup{colorlinks=true,urlcolor=darkred,linkcolor=darkred,citecolor=darkred}
\usepackage{verbatim}
\usepackage[letterpaper,margin=1in,headheight=15pt]{geometry}
\usepackage{mathpazo}
\usepackage{authblk}
\usepackage{empheq}

\usepackage[sorting=ydnt,bibstyle=authoryear-comp,labelyear=false,defernumbers=true,maxnames=20,firstinits=true, uniquename=init,dashed=false,backend=bibtex]{biblatex}

\DeclareNameAlias{sortname}{first-last}

\DeclareFieldFormat{url}{\url{#1}}
\DeclareFieldFormat[article]{pages}{#1}
\DeclareFieldFormat[inproceedings]{pages}{\lowercase{pp.}#1}
\DeclareFieldFormat[incollection]{pages}{\lowercase{pp.}#1}
\DeclareFieldFormat[article]{volume}{\textbf{#1}}
\DeclareFieldFormat[article]{number}{(#1)}
\DeclareFieldFormat[article]{title}{\MakeCapital{#1}}
\DeclareFieldFormat[inproceedings]{title}{#1}
\DeclareFieldFormat{shorthandwidth}{#1}

\DeclareBibliographyDriver{article}{%
  \usebibmacro{bibindex}%
  \usebibmacro{begentry}%
  \usebibmacro{author/editor}%
  \setunit{\labelnamepunct}\newblock
  \MakeSentenceCase{\usebibmacro{title}}%
  \newunit
  \printlist{language}%
  \newunit\newblock
  \usebibmacro{byauthor}%
  \newunit\newblock
  \usebibmacro{byeditor+others}%
  \newunit\newblock
  \printfield{version}%
  \newunit\newblock
  \usebibmacro{journal+issuetitle}%
  \newunit\newblock
  \printfield{note}%
  \setunit{\bibpagespunct}%
  \printfield{pages}
  \newunit\newblock
  \usebibmacro{eprint}
  \newunit\newblock
  \printfield{addendum}%
  \newunit\newblock
  \usebibmacro{pageref}%
  \usebibmacro{finentry}}

\renewbibmacro*{journal+issuetitle}{%
  \usebibmacro{journal}%
  \setunit*{\addspace}%
  \iffieldundef{series}
    {}
    {\newunit
     \printfield{series}%
     \setunit{\addspace}}%
  \printfield{volume}%
  \printfield{number}%
  \setunit{\addcomma\space}%
  \printfield{eid}%
  \setunit{\addspace}%
  \usebibmacro{issue+date}%
  \newunit\newblock
  \usebibmacro{issue}%
  \newunit}

\def\makebibcategory#1#2{\DeclareBibliographyCategory{#1}\defbibheading{#1}{\section*{#2}}}
\makebibcategory{books}{Books}
\makebibcategory{papers}{Refereed research papers}
\makebibcategory{chapters}{Book chapters}
\makebibcategory{conferences}{Papers in conference proceedings}
\makebibcategory{techreports}{Unpublished working papers}
\makebibcategory{bookreviews}{Book reviews}
\makebibcategory{editorials}{Editorials}
\makebibcategory{phd}{PhD thesis}
\makebibcategory{subpapers}{Submitted papers}
\makebibcategory{curpapers}{Current projects}

\renewcommand*{\bibitem}{\addtocounter{papers}{1}\item \mbox{}\hskip-0.85cm\hbox to 0.85cm{\hfill\arabic{papers}.~~}}
\defbibenvironment{bibliography}
{\list{}
  {\setlength{\leftmargin}{\bibhang}%
   \setlength{\itemsep}{\bibitemsep}%
   \setlength{\parsep}{\bibparsep}}}
{\endlist}
{\bibitem}

\newcounter{papers}
\newcounter{sumpapers}

\numberwithin{equation}{section}

\newcommand{\tops}{\texorpdfstring}

\newcommand{\fg}{{\mathfrak g}}

\newcommand{\cL}{\ensuremath{\mathcal L}}

\newcommand{\cM}{\ensuremath{\mathcal M}}

\newcommand{\R}{\ensuremath{\mathbb R}}
\newcommand{\C}{\ensuremath{\mathbb C}}
\newcommand{\PP}{\ensuremath{\mathbb P}}
\newcommand{\Z}{\ensuremath{\mathbb Z}}

\newcommand{\A}{\ensuremath{\mathbb A}}

\newcommand{\ctM}{\widetilde{\cM}}
\newcommand{\half}{\ensuremath{\frac{1}{2}}}

\newcommand{\N}{{\mathcal N}}

\newcommand{\hnabla}{{\widehat\nabla}}
\newcommand{\kahler}{K\"ahler\xspace}
\newcommand{\hk}{hyperk\"ahler\xspace}

\newcommand{\I}{{\mathrm i}}

\newcommand{\de}{\mathrm{d}}
\newcommand{\Sp}{{\mathrm{Sp}}}
\newcommand{\Spin}{{\mathrm{Spin}}}
\newcommand{\SU}{{\mathrm{SU}}}
\newcommand{\vol}{{\mathrm{vol}}}

\newcommand{\abs}[1]{\lvert#1\rvert}
\newcommand{\norm}[1]{\lVert#1\rVert}

\newcommand{\eps}{\epsilon}

\newcommand{\ti}[1]{\textit{#1}}

\DeclareMathOperator{\im}{Im}
\DeclareMathOperator{\re}{Re}

\bibliography{reduction-paper}

\begin{document}

\title{Hyperk\"ahler Sigma Model and Field Theory on Gibbons-Hawking Spaces}
\date{}
\author[1]{Anindya Dey}
\author[2]{Andrew Neitzke}
\affil[1]{Theory Group, University of Texas at Austin}
\affil[2]{Department of Mathematics, University of Texas at Austin}
\maketitle

{\abstract{We describe a novel deformation of the 3-dimensional sigma model with \hk target, which arises naturally from the compactification of a 4-dimensional $\mathcal{N}=2$ theory on a \hk circle bundle (Gibbons-Hawking space).  We derive the condition for which the deformed sigma model preserves 4 out of the 8 supercharges.  We also study the contribution from a NUT center to the sigma model path integral,
and find that supersymmetry implies it is a holomorphic section of a certain holomorphic line bundle over the \hk target.  We study explicitly the case where the original 4-dimensional theory is pure $U(1)$ super Yang-Mills, and show that the contribution from a NUT center in this case is simply the Jacobi theta function.
}}

\tableofcontents

\section{Introduction and main results}

It is a well-known principle that some aspects of quantum field theories become
easier to understand when the theories are compactified to lower dimensions. This principle was exploited in particular in \cite{Gaiotto:2008cd}, where the wall-crossing phenomenon
in $\N=2$ supersymmetric theories in four dimensions was studied by formulating the theories
on $S^1 \times \R^3$, with $S^1$ of fixed radius $R$.
A crucial input to that analysis was a good understanding of the constraints imposed by
supersymmetry \cite{Seiberg:1996nz,AlvarezGaume:1981hm}:  
they say that the IR Lagrangian of the compactified theory
is (around a generic point of its
moduli space) a sigma model into a \hk manifold $\cM[R]$.
The metric of $\cM[R]$ typically depends in a highly nontrivial way on the parameter $R$,
reflecting the fact that quantum corrections due to BPS particles of mass $M$ scale as $e^{-M R}$.

In this paper we consider a different but related problem:  we begin again with an $\N=2$ theory in
four dimensions, but rather than studying it on $S^1 \times \R^3$, we take our spacetime
to be a circle fibration over $\R^3$, with isolated degenerate fibers.  Generically such a
compactification would not preserve any supersymmetry, at least without some modification
of the theory; however, we consider the special case where the spacetime $X$ is actually \hk
(a Gibbons-Hawking space).  Thus $X$ has metric locally of the form
\begin{equation} \label{gh-metric}
\de s^2 = V \de{\vec{x}}^2 + \frac{R^2}{V}(\de\chi - B)^2
\end{equation}
where $\vec{x}$ is a coordinate in $\R^3$, $V(\vec x)$ a function on $\R^3$ (with singularities)
and $B$ a 1-form on $\R^3$.  (More globally $\de\chi - B$ is a connection
form in the circle bundle whose fiber coordinate is $\chi \in [0,4\pi]$.)
We take $V \to 1$ as $\vec x \to \infty$, so $R$ gives the asymptotic radius of compactification.
The \hk condition says that
\begin{equation} \label{gh-condition}
\star \de V = R \, \de B.
\end{equation}
Such an $X$ has holonomy $SU(2)$ rather than the generic $SU(2) \times SU(2) / \Z_2$, and
this reduced holonomy admits 4 covariantly constant spinors.  Thus the
resulting theory should have 4 supercharges.

\subsection*{A deformed \hk sigma model}

The first main question we address in this paper is:
what could the resulting theory look like from the three-dimensional point of view, after
reducing on the circle fiber?
Evidently it should be a deformation of the standard \hk sigma model,
which depends on the data of $V$ and $B$, which reduces to the original
model when $V$ is constant and $B = 0$, and which preserves $4$ supercharges
when \eqref{gh-condition} is satisfied.\footnote{We are \ti{not} saying that the theory
when \eqref{gh-condition} is satisfied
has $\N=2$ supersymmetry in three dimensions; it hardly could, since the function $V(\vec x)$
breaks the translation symmetry in $\R^3$.}

In \S\ref{sec:deformed-sigma-model} below we present a candidate form for such a deformation:
for the Lagrangian see \eqref{genhksigma}.
Our deformation involves some interesting geometry, which we now briefly describe.

\begin{itemize}
\item
First, the Lagrangian involves a one-parameter \ti{family} of \hk spaces.
More precisely, letting $\ctM$ denote the total space of this family, the Lagrangian
involves a bilinear form $g$ on the tangent space $T \ctM$, which restricts on each
fiber to a \hk metric.
The appearance of $\ctM$ might have been expected given the four-dimensional
origin of the model:  the spacetime metric \eqref{gh-metric} says that at different points
of $\R^3$ we should see different effective radii.

Being \hk, the fibers of $\ctM$ carry a $\C\PP^1$ worth of complex structures.
In the usual \hk sigma model, all these would be on the same footing, but in the deformed
model one of them is preferred.

\item
Second, the Lagrangian involves one extra
coupling, of the schematic form
\begin{equation} \label{new-cs-term}
 \frac{1}{8\pi} \int_{\R^3} \de B \wedge \varphi^* A
\end{equation}
where $A$ represents a $U(1)$ connection in a line bundle $\cL$ over the family $\ctM$, and $\varphi^* A$ is its
pullback to $\R^3$ via the sigma model field $\varphi$.
(We ignore for a moment the global topological issues involved in writing down \eqref{new-cs-term}.)

\item
We study the conditions under which the deformed theory preserves $4$ supercharges, and find
the following interesting consequence.  The family of manifolds $\ctM$ carries a preferred
torsion-free connection
$\hnabla$ in the tangent bundle, which preserves the preferred complex structures on the fibers
$\cM[\varphi^0]$, and agrees with the Levi-Civita connection fiberwise.  Moreover,
the $\hnabla$-covariant derivative of the bilinear form $g$ is constrained in terms of $F$,
as expressed in \eqref{zeroshift1}-\eqref{zeroshift2} below.

\end{itemize}

\subsection*{Contributions from NUT centers}

The second main question we address is what happens around places where
the function $V$ in \eqref{gh-metric} becomes singular, with
\begin{equation}
V \sim \frac{R}{r}.
\end{equation}
At these points (sometimes called ``NUT centers'')
the circle fiber shrinks to zero size, and the dimensional reduction procedure needs
to be modified.  We deal with this by cutting out a small neighborhood of
each NUT center; thus the physics very near the NUT center is ``integrated out'' and replaced
by some effective interaction for the fields on the boundary $S^3$.  After compactification
to three dimensions the boundary is an $S^2$.   At the lowest order in the derivative expansion,
the boundary interaction is roughly a function $Q(\varphi)$ of the value of $\varphi$ along
this $S^2$, i.e. a function
\begin{equation}
 Q: \ctM \to \R.
\end{equation}
In \S\ref{boundarySUSY} we work out the constraints imposed by supersymmetry on this kind of
boundary interaction.  The answer depends on a topological invariant of the situation, namely
the degree $k$ of the circle bundle over the boundary $S^2$,\footnote{There is an unconventional
factor of $2$ in \eqref{degree}, which will recur in various other equations in this paper;
this factor comes ultimately from our convention that the coordinate $\chi$ has period $4\pi$
rather than $2\pi$.}
\begin{equation} \label{degree}
 k = \frac{1}{4\pi} \int_{S^2} G.
\end{equation}
We find (with respect to
the preserved complex structure on the fiber $\cM[\varphi^0]$)
\begin{equation} \label{pert-holomorphy}
 \bar\partial Q + k A^{(0,1)} = 0.
\end{equation}
In particular, for a boundary component around a NUT center
the circle bundle is the Hopf fibration $S^3 \to S^2$, which has
degree $k=-1$, so we get
\begin{equation}
 \bar\partial Q - A^{(0,1)} = 0.
\end{equation}

Geometrically,
in order for \eqref{pert-holomorphy} to make global sense, $Q$ should not be quite
a function --- rather, $e^Q$ should be a section of $\cL^k$, where $\cL$ is the line
bundle introduced above (on which $A$ is a connection).
\eqref{pert-holomorphy} then says that $e^Q$ is actually a \ti{holomorphic} section of $\cL$,
with respect to a holomorphic structure on $\cL$ determined by $A^{(0,1)}$.
This is a very strong constraint on $Q$, since any two such sections differ by a global holomorphic
function, in the preserved complex structure, and $\cM$ has rather few global holomorphic functions.

\subsection*{Topological issues}

Now let us return to some topological issues we ignored above.
The term \eqref{new-cs-term} is a bit subtle, since as written it only makes sense when $B$ is
a connection in a \ti{trivial} bundle.  By integration by parts we could try to move the problem
over to $A$, but in the examples which occur in nature, both $A$ and $B$
are actually connections in topologically nontrivial bundles.
The problem is similar to the problems one meets in defining the $U(1)$ Chern-Simons
interaction,
\begin{equation}
 \int \de A \wedge A,
\end{equation}
except that in \eqref{new-cs-term} we have two line bundles with connection involved rather than one,
and one of the bundles arises by pullback from the space $\ctM$.

As with the usual Chern-Simons story, even though the action does not make global sense,
the \ti{exponentiated} action may still be well defined, provided that the coefficient of
the problematic term is properly quantized.  That is the case here, so
\eqref{new-cs-term} is not a problem, at least on a compact three-manifold.

However, as we have discussed above, we will want to consider the effective three-dimensional theory
on manifolds with boundary (obtained by cutting out the NUT centers.)
In this case we meet a further subtlety in defining \eqref{new-cs-term}, again well known from
Chern-Simons theory:  even the exponential of \eqref{new-cs-term} is generally not well defined as a complex number.
Rather it must be interpreted as an element of a certain complex ``Chern-Simons line'' depending on the boundary
value of $\varphi$.  How are we to square this
with the expectation from the original four-dimensional theory, where the exponentiated action
seems to be a number in the usual sense?
The resolution is that there is also a contribution from the boundary term
$Q(\varphi)$, and we recall that $e^Q$ is valued in the line $\cL$.
Thus, everything will be consistent if the Chern-Simons line where the exponential of
\eqref{new-cs-term} lives is precisely the dual line $\cL^*$.  This is indeed the case.

\subsection*{The simplest example}

In this paper we study in some detail one concrete example of this general story, the simplest possible one:
pure $\N=2$ theory with gauge group $U(1)$.
By direct computation we find that the compactified theory is indeed described by
a \hk sigma model deformed in accordance with our general recipe.
The space $\ctM$ in this case is a 5-manifold, fibered over the real line parameterized
by $\varphi^0 = \frac{V}{R}$.  Writing the fibers in terms of their preserved complex structure, we have
\begin{equation}
 \cM[\varphi^0] \simeq \C \times T^2,
\end{equation}
where the $T^2$ factor has complex modulus
$\tau$ given by the complexified gauge coupling.  (This space is the most trivial example of a Seiberg-Witten
integrable system.)

The line bundle $\cL$ has nontrivial topology over each $T^2$ fiber; more precisely, the possible
topologies are classified by an integer (degree), and $\cL$ has degree $1$.
Holomorphically it is the famous ``theta line bundle,'' of which the theta function
is a holomorphic section.  Not surprisingly, then, the boundary term $e^Q$ turns out to be
a theta function:
\begin{equation}
 e^Q = \theta(\tau, z).
\end{equation}
As we have explained, this form is essentially dictated by supersymmetry, but
we can also understand the appearance of this theta function directly:
it arises from a sum over smooth $U(1)$ instantons
supported near the NUT center.  We work this out in \S\ref{NUToperator}.

\subsection*{Discussion and connections}

\begin{itemize}

\item Our results in this paper fit in well with the observation in \cite{Witten:2009at} that in compactifications of the
$(2,0)$ SCFT from 6 to 5 dimensions on a circle bundle one gets a 5-dimensional supersymmetric Yang-Mills theory, coupled to a 2-dimensional WZW model at each codimension-3 locus where the circle fiber degenerates.
Indeed, upon further compactification on a Riemann surface $C$, this suggests that in the class $S$ theory
$S[\fg,C]$, the contributions from NUT centers should be something like the partition function of the WZW model
with group $G$.  The particular case which we consider here is essentially the case $G = U(1)$ and $C = T^2$, for which the WZW partition function is an ordinary theta function, indeed matching what we find for
$e^Q$.  It would be very interesting to understand how to recover the ``nonabelian theta functions''
by analogous computations in interacting four-dimensional field theories.

\item The problem we consider bears some similarity to one described in \cite{Cordova:2013bea}, where the authors consider the dimensional reduction of the $(2,0)$
superconformal field theory from six to five dimensions on a circle bundle, and
obtain a deformed version of five-dimensional super Yang-Mills.
It would be interesting to know whether the two constructions fit inside a common framework.

\item The original motivation for this work was the results of \cite{Neitzke:2011za,Alexandrov:2011ac},
where it was found that the moduli space $\cM[R]$ which appears in compactification
of an $\N=2$ theory on $S^1$ carries a natural line bundle $V$
which admits a \ti{hyperholomorphic} structure (see also \cite{MR2394039,Hitchin2012} for mathematical
accounts of the same bundle).  In particular, it was conjectured in \cite{Neitzke:2011za} that
the contribution to the 3d effective theory from a NUT center would be a holomorphic section of $V$.

In this paper our formalism is slightly different from that envisaged in \cite{Neitzke:2011za}.
Our deformed sigma model involves a family of moduli spaces $\cM[\varphi^0]$
which have \ti{a priori} nothing to do with the spaces $\cM[R]$, since the theory
on a general Gibbons-Hawking space has \ti{a priori} nothing to do with the theory
on $\R^3 \times S^1$.
Still, the two models can be related to one another,
at least when the original four-dimensional theory is
conformally invariant.  Indeed, by a local conformal transformation
we can change the Gibbons-Hawking metric \eqref{gh-metric} to
\begin{equation} \label{gh-metric2}
\de s^2 = \de{\vec{x}}^2 + \frac{R^2}{V^2}(\de\chi - B)^2
\end{equation}
so that when $V$ is slowly varying, the theory looks locally like a compactification on $\R^3 \times S^1$
for which the radius of $S^1$ is $R/V$.

Using this relation we can try to compare our results with the expectations from \cite{Neitzke:2011za}.
In the example considered in \S\ref{GH-U(1)ex} and \S\ref{NUToperator}, we indeed find that $e^Q$ is a holomorphic section of a holomorphic bundle $\cL$ which is holomorphically equivalent to $V$.

\end{itemize}

\subsection*{Acknowledgements}

We thank Jacques Distler, Dan Freed, Daniel Jafferis, Greg Moore and Boris Pioline for useful discussions. The research of AD is supported by the National Science Foundation under grant numbers PHY-1316033 and PHY-0969020. The research of AN is supported by NSF grant 1151693.

\section{The deformed \hk sigma model} \label{sec:deformed-sigma-model}

\subsection{Fields of the undeformed model}

The standard \hk sigma model \cite{Bagger:1983tt} in three dimensions involves a single \hk target space $\cM$.
Let the dimension of $\cM$ be $4r$.
Recall that the complexified tangent bundle of $\cM$ admits a decomposition
\begin{equation} \label{eq:tangent-decomp}
 T_\C \cM = H \otimes E,
\end{equation}
invariant under the Levi-Civita connection.
Here $E$ is an $\Sp(r)$ bundle of dimension $2r$,
and $H$ is a trivial $\Sp(1)$ bundle of dimension $2$.

The sigma model fields are
\begin{align}
 \varphi &: \R^3 \to \cM, \\
 \psi,\bar{\psi} &\in \Gamma(S \otimes \varphi^* E),
\end{align}
where $S$ is the (complex, two-dimensional) spinor representation of $\Spin(3) \simeq \SU(2)$.

In this note, we use unprimed uppercase Latin letters for $Sp(1)$ indices and primed uppercase Latin letters for $Sp(r)$ indices.  Spinor indices will be denoted by lowercase Greek letters while lowercase Latin letters are used to label the local coordinates on the \hk space.
Thus in components the fields would be written $\varphi^i$ ($i = 1, \dots, 4r$) and $\psi_{A'\alpha}, \bar{\psi}^{A'}_{\alpha}$
($A' = 1, \dots, r$; $\alpha = 1, 2$.)

\subsection{Data for the deformed model}

Our deformed model involves not a single \hk space but a family of them, parameterized by a new
scalar which we will call $\varphi^0$.  Let $\ctM$ denote the full family, which is $(4r+1)$-dimensional,
with local coordinates $\varphi^i$ ($i = 0, \dots, 4r$).

We emphasize that $\varphi^0$ is \ti{not} a field in the deformed sigma model:  rather it will be
a fixed background function on $\R^3$.  We require that $\varphi^0$ is harmonic
on $\R^3$ (perhaps with singularities),
and moreover that there is a line bundle over $\R^3$ (away from the singularities of $\varphi^0$)
with connection $B$ and curvature $G$,
such that \begin{equation}
G = \star \de \varphi^0.
\end{equation}

$\ctM$ carries a bilinear form $g$, which
restricts to the \hk metric on each fiber $\cM[\varphi^0]$, but which need not be nondegenerate
on the whole family.

$\ctM$ also carries a line bundle $\cL$ with connection.  This is one of the key
new ingredients in our deformed model,
with no direct analogue in the ordinary \hk sigma model.\footnote{Thus we have \ti{two} line bundles with connection in the story, one over $\R^3$ with connection $B_\mu$
($\mu = 1, 2, 3$), the other over
$\ctM$ with connection $A_i$ ($i = 0, \dots, 4r$); the two should not be confused.}
Locally we may trivialize this line bundle and thus represent the connection by a $1$-form $A$ on $\ctM$,
whose curvature is $F$.  We work in conventions where $A$ and $F$ are purely imaginary.

Being \hk, each fiber $\cM[\varphi^0]$ carries a family of complex structures
parameterized by lines in $H$, i.e. points of the projective space
$\PP(H)$.  In the deformed sigma model one point $c \in \PP(H)$ will be distinguished,
corresponding to a preferred complex structure on each fiber $\cM[\varphi^0]$.
By a rotation of the complex structures on $\ctM$ we can always
choose $c^A = (0,1)$.
In what follows we will always make this choice.

Thus we have two $2r$-dimensional distributions $T^{1,0}$, $T^{0,1}$ on $\ctM$,
consisting respectively of $(1,0)$ or $(0,1)$ vectors tangent to the fibers.
They induce the structure of a (Levi-flat) CR manifold on $\ctM$.

\subsection{Hyperk\"ahler identities and their extensions}\label{exthk}

The supersymmetry of the \hk sigma model depends on certain identities which are part of the standard
story of \hk geometry.  In our deformed \hk sigma model we will need a slightly different geometric
structure, which involves some extensions of these identities.
Here we briefly review the relevant identities and state the requisite
extensions.

One of the fundamental objects which enters the \hk sigma model is the isomorphism
$e: T_\C \cM \to H \otimes E$, represented in local coordinates as $e_{iEE'}$.
This isomorphism takes the Levi-Civita connection in $T_\C \cM$
to an $\Sp(r)$-connection in $E$,
which we write in local coordinates as $q_{jB'}^{A'}$.  This statement is expressed
by the identity
\begin{equation} \label{eq:e-cc1}
\partial_j e_{iEE'} - q_{jE'}^{A'} e_{iEA'} = \Gamma_{ji}^k e_{kEE'}  \, \text{ for } \, i,j \neq 0.
\end{equation}
In the standard \hk sigma model, \eqref{eq:e-cc1}, combined with the standard formula
for $\Gamma_{ji}^k$ in terms of $g$,
ensures that the 1-fermion terms in the SUSY variation of the sigma model action vanish.

In our deformed sigma model, the bundles $E$ and $H$ will be extended over the
full $\ctM$, as will the $\Sp(r)$-connection $q$ in $E$; moreover the isomorphism $e$ will be
extended to a surjection $e: T_\C \ctM \to H \otimes E$. We will also extend the Levi-Civita connection
to a connection $\nabla$ in the full $T_\C \ctM$, given in coordinates by symbols $\Gamma_{ji}^k$
where now $i,j,k$ run from $0$ to $4r$.
Finally, we will define a shifted version of $q$, of the form
\begin{equation}
\tilde{q}_{jE'}^{A'} = q_{j E'}^{A'} + f(\varphi^0) \delta^0_j \delta_{E'}^{A'},
\end{equation}
for some function $f$.
The key identity \eqref{eq:e-cc1} will then be extended to
\begin{empheq}[box=\fbox]{align}
\begin{split}
\partial_j e_{iEE'} - \tilde{q}_{jE'}^{A'} e_{iEA'} = \Gamma_{ji}^k e_{kEE'} \label{eq:e-cc2}
\end{split}
\end{empheq}
where now all indices run from $0$ to $4r$.
For $i,j \neq 0$, we would like \eqref{eq:e-cc2} to reduce to \eqref{eq:e-cc1}; thus we will require
$\Gamma_{ji}^0=0$ for $i,j \neq 0$.  In addition, we choose $\Gamma^0_{i0} = 0$, $\Gamma^0_{0i}=0$.
Thus altogether
\begin{equation} \label{gamma0}
\Gamma_{ji}^0 = 0.
\end{equation}
More invariantly, this says that the extended connection $\nabla$ preserves the distribution of
vertical tangent vectors on $\ctM$ and that $\nabla (\partial_0)$ is also vertical.
The remaining components $\Gamma_{0i}^k$ and $\Gamma_{i0}^k$ of $\nabla$
are determined by requiring \eqref{eq:e-cc2}.  Thus $\nabla$ is completely determined
once the extended $e$ and $q$ and the function $f$ are given.

As we will show in \S\ref{1-fermionSUSY} below,
vanishing of the 1-fermion terms in the deformed sigma model action leads to a condition on
$\nabla$: for $E=2$, we will need
\begin{empheq}[box=\fbox]{gather}
e^l_{EE'} \left( \partial_i g_{jl} + \partial_j g_{il} - \partial_l g_{ij} - ({\Gamma}_{ij}^k+{\Gamma}_{ji}^k) g_{kl} + \delta_{j}^0 F_{il} + \delta_i^0 F_{jl} \right) = 0, \label{covg} \\
e_{kEE'} (\Gamma^k_{ji} - \Gamma^k_{ij}) = 0. \label{torsionfree}
\end{empheq}
The equations \eqref{covg}-\eqref{torsionfree} constitute one of the main results of this paper.

To assess the geometric content of \eqref{covg}-\eqref{torsionfree} it is convenient to look
not at $\nabla$ but rather at a closely related
connection $\hnabla$.  $\hnabla$ is characterized by the requirements that it is a \ti{real} connection
and that, for any vector fields $Y$ and $Z$,
\begin{equation} \label{hnabla-cond}
 \nabla_Y Z - \hnabla_Y Z \in T^{0,1}.
\end{equation}
In local coordinates this says that the coefficients
$\widehat\Gamma_{ji}^k$ agree with $\Gamma_{ji}^k$ whenever $k$ is a holomorphic direction
or the $\varphi^0$ direction, while the $\widehat\Gamma_{ji}^k$ for $k$ an antiholomorphic direction
are determined by the requirement that $\hnabla$ is real.

Then \eqref{torsionfree} says that $\hnabla$
is torsion-free.  Indeed, requiring \eqref{torsionfree} for $E=2$ along with \eqref{gamma0}
implies that the torsion
of $\nabla$ is valued in $T^{0,1}$, i.e.
\begin{equation}
T_\nabla(X,Y) = \nabla_X Y - \nabla_Y X - [X,Y] \in T^{0,1}
\end{equation}
but then using \eqref{hnabla-cond} we have
\begin{equation}
T_{\hnabla}(X,Y) = \hnabla_X Y - \hnabla_Y X - [X,Y] \in T^{0,1},
\end{equation}
and if $X$, $Y$ are real then $T_\hnabla(X,Y)$ is also real, but the only
way that something real can lie in $T^{0,1}$ is if it actually vanishes, i.e.
$T_\hnabla(X,Y) = 0$.

Next, we consider \eqref{covg}.
As $E'$ varies, with $E=2$, the vector field $e^l_{EE'} \partial_l$ runs over a basis for
$T^{0,1}$; thus the quantity in parentheses in \eqref{covg} vanishes
whenever $l$ is an antiholomorphic direction, for any $i,j$.
Moreover, using the fact that $\Gamma^0_{ij} = 0$ and $g$ is Hermitian on each fiber of $\ctM$,
the only terms which contribute in \eqref{covg} when $l$ is antiholomorphic will be those with $k$
holomorphic; for these moreover we have $\hat\Gamma_{ij}^k = \Gamma_{ij}^k$, so
finally we get for $i,j \neq 0$ and $l$ antiholomorphic
\begin{equation} \label{lhol}
\partial_i g_{jl} + \partial_j g_{il} - \partial_{l} g_{ij} - 2 \hat\Gamma_{ij}^k g_{kl} = 0.
\end{equation}
Now since both $g$ and $\hat\Gamma$ are real, we may simply take the complex conjugate of \eqref{lhol}
to get the same equation with $l$ holomorphic.  Thus \eqref{lhol} holds for all $i,j,l \neq 0$.
This is just the standard formula for the Levi-Civita connection.  Thus
\eqref{covg} requires that $\hnabla$ agrees with the Levi-Civita connection on each fiber
of $\ctM$.

Now consider what \eqref{covg} says if we take $i = 0$, $j \neq 0$, and $l$
antiholomorphic:
\begin{equation}
\partial_0 g_{jl} + \partial_j g_{0l} - \partial_l g_{0j} - 2 {\hat\Gamma}_{0j}^k g_{kl} + F_{jl} = 0,
\end{equation}
which we could also write as
\begin{equation} \label{zeroshift1}
\hnabla_0 g_{jl} + \hnabla_j g_{0l} - \hnabla_l g_{0j} + F_{jl} = 0.
\end{equation}
If $i = 0$, $j = 0$ and $l$ is antiholomorphic then we get similarly
\begin{equation} \label{zeroshift2}
2 \hnabla_0 g_{0l} - \hnabla_l g_{00} + 2 F_{0l} = 0.
\end{equation}
The equations \eqref{zeroshift1}-\eqref{zeroshift2} are expressing the constraint
imposed by supersymmetry on the $\varphi^0$-dependence of the bilinear form $g$.
It would be interesting to understand better their intrinsic geometric meaning.
Note that if $g$ is fiberwise covariantly constant then \eqref{zeroshift1}
reduces to the pleasant form $\hnabla_0 g_{jl} + F_{jl} = 0$; this is indeed true in
the simple example we consider in \S\ref{GH-U(1)ex}, but we do not know whether it will be the
case generally.

Another identity coming from the special form of the curvature for a \hk manifold is
\begin{equation} \label{eq:hk-curvature1}
(\partial_j q^{A'}_{i E'} - \partial_i q_{j E'}^{A'} + q_{j B'}^{A'} q^{B'}_{i E'} - q^{A'}_{i B'} q_{j E'}^{B'}) e^j_{EF'} = \Omega^{A'}_{F'B'E'} e_{iE}^{B'}  \, \text{ for } \, i \neq 0.
\end{equation}
The 3-fermion terms in the SUSY variation of the standard \hk sigma model action vanish provided this identity is satisfied.  Vanishing of the 3-fermion terms in the deformed model
needs a simple extension of \eqref{eq:hk-curvature1}:  we simply require that the same equation
holds even for $i=0$, i.e.
\begin{equation} \label{eq:hk-curvature2}
\boxed{(\partial_j q^{A'}_{i E'} - \partial_i q_{j E'}^{A'} + q_{j B'}^{A'} q^{B'}_{i E'} - q^{A'}_{i B'} q_{j E'}^{B'}) e^j_{EF'} = \Omega^{A'}_{F'B'E'} e_{iE}^{B'}  \,  \, \forall i}
\end{equation}

Finally, if we define
\begin{equation}\label{bianchi0}
B_{i A'B'C'D'} = \partial_i \Omega_{A'B'C'D'} - q_{i A'}^{E'} \Omega_{E'B'C'D'} - q_{i B'}^{E'} \Omega_{A'E'C'D'} - q_{i C'}^{E'} \Omega_{A'B'E'D'} - q_{i D'}^{E'} \Omega_{A'B'C'E'},
\end{equation}
then for $i \neq 0$ we have the Bianchi identity,
\begin{equation}\label{bianchi1}
B_{i A' B' C' D'} e^i_{FF'} - B_{i F'B'C'D'} e^i_{F A'} = 0.
\end{equation}
The identity \eqref{bianchi1} ensures that 5-fermion terms in the SUSY variation of the
standard \hk sigma model action vanish.  The same identity will suffice
for the deformed model as well (said otherwise, the extension of \eqref{bianchi1}
to include $i = 0$ is automatically satisfied, since $e^0_{AA'} = 0$.)

\subsection{Action of the deformed model}\label{defaction-hk}

The action for our deformed sigma model is:
\begin{equation}
\begin{split}
4 \pi \mathcal{L} & = \half \partial_\mu \varphi^i \partial^\mu \varphi^j g_{ij} \\
&- \I \bar\psi_{\alpha A'} \gamma^{\mu \alpha}_\beta (\partial_\mu \psi^{A'\beta} + q^{A'}_{B' i} \partial_\mu \varphi^i \psi^{B' \beta}) \\
&+ \half \Omega_{A'B'C'D'} (\psi^{A'}_{\alpha} \gamma^{\mu\alpha}_\beta \bar\psi^{B'\beta}) (\psi^{C'}_\delta \gamma_{\mu \omega}^\delta \bar\psi^{D'\omega}) \\
&+ \half \eps_{\mu \nu \rho} G^{\mu\nu} \partial^\rho \varphi^i A_i.
\end{split} \label{genhksigma}
\end{equation}
In the special case where $G = 0$ and $\varphi^0$ is constant, this action reduces
to the undeformed \hk sigma model as written in \cite{Bagger:1983tt}.

The SUSY transformations are generated by fermionic parameters $\zeta^\alpha_E, \bar\zeta^E_\alpha$ as follows:
\begin{align}
\delta_\zeta \varphi^i &=  \psi^{E'\alpha} \bar\zeta^E_\alpha e^i_{EE'} + \bar\psi_{E'\alpha} \zeta^\alpha_E e^{i E E'}, \label{susyvar1}\\
\delta_\zeta \psi^{A'}_\alpha &= - \I \partial_\nu \varphi^i e_i^{EA'} \gamma_\alpha^{\nu \sigma} \zeta_{\sigma E} - q_{i E'}^{A'} \delta_\zeta \varphi^i \psi^{E'}_\alpha, \label{susyvar2}\\
\delta_\zeta \bar \psi^{\alpha}_{A'} &= -\I \partial_\nu \varphi^i e_{iEA'} \gamma_\sigma^{\nu \alpha} \bar\zeta^{\sigma E} + q_{i A'}^{E'} \delta_\zeta \varphi^i \bar\psi_{E'}^\alpha. \label{susyvar3}
\end{align}

In the undeformed sigma model there is an $8$-dimensional space of possible SUSY parameters
$(\zeta^\alpha_E$, $\bar\zeta^E_\alpha)$.
In our deformed model we typically have $\delta_\zeta \cL = 0$ only for $(\zeta^\alpha_E$, $\bar\zeta^E_\alpha)$ obeying
\begin{equation}
 c^A \zeta^\alpha_A = 0, \qquad c_A \bar\zeta^A_\alpha = 0.
\end{equation}
These equations reduce the supersymmetries from $8$ to $4$.\footnote{Note that the supersymmetries
preserved are \ti{not} those which
correspond to an ordinary \kahler sigma model \cite{Bagger:1982fn} into $\cM$ with its preferred
complex structure, although that model also has $4$ supercharges.
With our preserved supersymmetries we always have $\zeta^\alpha_E \bar\zeta^E_\beta = 0$, reflecting the
fact that all translations are broken (as they must be, since the background field $\varphi^0$ generally
has no translation symmetry.)}
If we choose $c^A = (0,1)$ as mentioned above, the supersymmetries
are generated by $\zeta^\alpha_1$ and $\bar\zeta^2_\alpha$.

The parameter $\zeta_E^\alpha$ is not a constant spinor:  it may depend on position through $\varphi^0$.
We will require
\begin{equation}
\begin{split}
&\partial_{\mu} \zeta^{\alpha}_E + f(\varphi^0) \partial_{\mu} \varphi^0\zeta^{\alpha}_E=0, \\
&\partial_{\mu} \bar{\zeta}^{\alpha}_E + f(\varphi^0) \partial_{\mu} \varphi^0\bar{\zeta}^{\alpha}_E=0,
\end{split} \label{SUSYparHK}
\end{equation}
where $f(\varphi^0)$ is a function of the background scalar $\varphi^0$ only. Finally, since $\varphi^0$ is a background field we should have $\delta_\zeta \varphi^0 = 0$: thus we require $e^0_{EE'} = 0$.\\

For our purposes it will not be necessary to write the explicit reality condition on the spinors or the SUSY parameters.
We will simply treat the barred and the unbarred spinors as independent 2-component complex spinors.
We use the following conventions for contracting spinors and gamma matrices in three dimensions:
\begin{equation}
\eps^{12} = 1 = \eps_{21}, \quad \psi^\alpha = \psi_\beta \eps^{\beta\alpha}, \quad \psi_\alpha = \psi^\beta \eps_{\beta \alpha},
\end{equation}
\begin{equation}
\gamma^{\mu \alpha \beta} = \gamma^{\mu \alpha}_\sigma \eps^{\sigma \beta} = \gamma^{\mu \beta \alpha}, \qquad \psi^\alpha \lambda_\alpha = - \psi_\alpha \lambda^\alpha
\end{equation}
The $Sp(r)$ indices are raised and lowered by antisymmetric matrices $\epsilon^{A'B'}$ and $\epsilon_{A'B'}$ which are covariantly constant. The $Sp(1)$ indices are similarly raised/lowered by constant antisymmetric matrices $\epsilon^{AB}$ and $\epsilon_{AB}$.

\subsection{SUSY verification:  1-fermion terms}\label{1-fermionSUSY}

The SUSY variation of the deformed sigma model action can be decomposed into terms involving 1, 3, and 5 fermions.  The 3 and 5-fermion terms have exactly the same structure as in the undeformed \hk sigma model, and the vanishing of these terms works out in the standard fashion; thus we relegate
their discussion to Appendix \ref{susy35}. The new contributions come exclusively from the 1-fermion terms;
vanishing of these terms leads to an interesting new equation, as we shall now describe.

The terms with a single fermion in the variation of the Lagrangian are:
\begin{subequations} \label{eq:psi}
\begin{align}
4 \pi \delta \mathcal{L} =
&+ \partial_\mu ( \psi^{E'\alpha} \bar\zeta^E_\alpha e^i_{EE'}) \partial^\mu \varphi^j g_{ij} \\
&+ \half \partial_\mu \varphi^i \partial^\mu \varphi^j \partial_k g_{ij}  \psi^{E'\alpha} \bar\zeta^E_\alpha e^k_{EE'} \\
&-  \partial_\nu \varphi^i e_{iEA'} \gamma_{\sigma\alpha}^{\nu} \bar\zeta^{\sigma E} \gamma^{\mu \alpha}_\beta (\partial_\mu \psi^{A'\beta} + q^{A'}_{B' j} \partial_\mu \varphi^j \psi^{B' \beta}) \\
&+ \half \eps_{\mu \nu \rho} G^{\mu\nu} \partial^\rho ( \psi^{E'\alpha} \bar\zeta^E_\alpha e^i_{EE'}) A_i \\
&+ \half \eps_{\mu \nu \rho} G^{\mu\nu} \partial^\rho \varphi^i \partial_k A_i  \psi^{E'\alpha} \bar\zeta^E_\alpha e^k_{EE'}
\end{align}
\end{subequations}

Note that we have only retained terms linear in $\bar\zeta^E_\alpha$ in the variation. Terms linear in $\zeta^\alpha_E$ can be formally obtained by complex conjugation. Integrating by parts to put all derivatives on $\varphi$
gives
\begin{subequations}
\begin{align}
4 \pi \delta \mathcal{L} =
&- \psi^{E'\alpha} \bar\zeta^E_\alpha e^i_{EE'}  \partial_\mu (\partial^\mu \varphi^j g_{ij}) \\
&+ \half \partial_\mu \varphi^i \partial^\mu \varphi^j \partial_k g_{ij} \psi^{E'\alpha} \bar\zeta^E_\alpha e^k_{EE'} \\
& +  \partial_\mu(\partial_\nu \varphi^i e_{iEA'} \gamma_{\sigma\alpha}^{\nu} \bar\zeta^{\sigma E} \gamma^{\mu \alpha}_\beta)  \psi^{A'\beta} \\
&- \partial_\nu \varphi^i e_{iEA'} \gamma_{\sigma\alpha}^{\nu} \bar\zeta^{\sigma E} \gamma^{\mu \alpha}_\beta q^{A'}_{B' j} \partial_\mu \varphi^j \psi^{B' \beta} \\
& - \half \eps_{\mu \nu \rho} \partial^\rho (G^{\mu\nu} A_i) \psi^{E'\alpha} \bar\zeta^E_\alpha e^i_{EE'} \\
&+ \half \eps_{\mu \nu \rho} G^{\mu\nu} \partial^\rho \varphi^i \partial_k A_i \psi^{E'\alpha} \bar\zeta^E_\alpha e^k_{EE'}
\end{align}
\end{subequations}
Next, we rewrite the variation as  $\delta \mathcal{L} = \bar\zeta^E_\alpha \delta \mathcal{L}^\alpha_E$ and $\delta \mathcal{L}^\alpha_E = \delta \mathcal{L}^{\alpha}_{\beta EE'} \psi^{E'\beta}$.  Using the relations  $\half \{\gamma^\nu, \gamma^\mu\} = g^{\mu \nu}$ and $e^i_{EE'} g_{ij} = e_{jEE'}$ as well as $\partial_{\mu} \bar{\zeta}^{\alpha}_E + f(\varphi^0) \partial_{\mu} \varphi^0\bar{\zeta}^{\alpha}_E=0$, the above expression may be reduced to
\begin{subequations}
\begin{align}
4 \pi \delta \mathcal{L}^\alpha_{\beta EE'} =
&+ \partial^\mu \varphi^j \partial_\mu \varphi^i \partial_i g_{kj} e^k_{EE'}  \delta^\alpha_\beta \\
&- \half \partial_\mu \varphi^i \partial^\mu \varphi^j \partial_k g_{ij} e^k_{EE'} \delta^\alpha_\beta \\
&- \partial_\nu \varphi^i \partial_\mu \varphi^j \partial_j e_{iEE'} (\gamma^\nu\gamma^\mu)^{\alpha}_\beta \\
&+f(\varphi^0) \partial_\mu \varphi^0 \partial_\nu \varphi^i e_{iEE'} (\gamma^\nu\gamma^\mu)^{\alpha}_\beta \\
&+ \partial_\nu \varphi^i \partial_\mu \varphi^j e_{iEA'} q^{A'}_{E' j} (\gamma^\nu\gamma^\mu)^{\alpha}_\beta \\
&+ \half \delta^\alpha_\beta \eps_{\mu \nu \rho} {\partial^\rho} (G^{\mu\nu} A_i) e^i_{EE'} \\
&- \half \delta^\alpha_\beta \eps_{\mu \nu \rho} G^{\mu\nu} \partial^\rho \varphi^i \partial_k A_i e^k_{EE'}
\end{align}
\end{subequations}
Now recall \eqref{eq:e-cc2}:
\begin{align}
&\partial_j e_{iEE'} - \tilde{q}_{jE'}^{A'} e_{iEA'} = \Gamma_{ji}^k e_{kEE'}, \nonumber \\
&\tilde{q}_{jE'}^{A'}=q_{j E'}^{A'}+f(\varphi^0) \delta^0_j \delta_{E'}^{A'}. \nonumber
\end{align}
Using this we can eliminate the $q_j$ in favor of the relevant components $\Gamma_{ij}^k$.  We split the result in the form $\delta \mathcal{L}^\alpha_{\beta EE'} = \delta^\alpha_\beta \delta \mathcal{L}_{EE'} + (\gamma^{\rho})^\alpha_\beta \delta \mathcal{L}_{\rho EE'}$,
where
\begin{subequations}
\begin{align}
4 \pi\delta \mathcal{L}_{EE'} =
&+ \partial^\mu \varphi^j \partial_\mu \varphi^i \partial_i g_{kj} e^k_{EE'} \\
&- \half \partial_\mu \varphi^i \partial^\mu \varphi^j \partial_k g_{ij} e^k_{EE'} \\
&- \half \partial_\mu \varphi^i \partial^\mu \varphi^j (\Gamma_{ji}^k+\Gamma_{ij}^k) e_{kEE'} \\
&+ \half \eps_{\mu \nu \rho} {\partial^\rho} (G^{\mu\nu} A_i) e^i_{EE'} \\
&- \half \eps_{\mu \nu \rho} G^{\mu\nu} \partial^\rho \varphi^i \partial_k A_i e^k_{EE'}
\end{align}
\end{subequations}
\begin{align}
4 \pi \delta \mathcal{L}_{\rho EE'} = -\I \epsilon_{\mu\nu\rho} \partial_{\mu} \varphi^i \partial_{\nu} \varphi^j e_{kEE'} (\Gamma^k_{ji} -\Gamma^k_{ij})
\end{align}
Recall that $\Gamma^k_{ij}$ is not necessarily symmetric in its lower indices (the connection $\nabla$
may have torsion.)
$\delta \mathcal{L}_{EE'}$ involves only the symmetric part of $\Gamma^k_{ij}$,
while the antisymmetric part is contained in $\delta \mathcal{L}_{\rho EE'}$.
Rearranging the terms in $\delta \mathcal{L}_{EE'}$, we have
\begin{align}
4 \pi \delta \mathcal{L}_{EE'} = & \half \partial^\mu \varphi^i \partial_\mu \varphi^j e^l_{EE'} \left(  \partial_i g_{jl} + \partial_j g_{il} - \partial_l g_{ij} - ({\Gamma}_{ij}^k+{\Gamma}_{ji}^k) g_{kl} + \delta_{j}^0 F_{il} + \delta_i^0 F_{jl} \right),
\end{align}
where we have used $\de G = 0$ and $\half \eps_{\mu \nu \rho} G^{\mu\nu} = \partial_\rho \varphi^0$ to consolidate the last two terms into $\partial_\rho \varphi^0 {\partial^\rho} \varphi^i F_{ik} e^k_{EE'}$.
Thus, finally, the condition for supersymmetry of the deformed \hk sigma model can be summarized as
\begin{align}
\delta \mathcal{L}_{EE'} = 0 & \implies e^l_{EE'} \left( \partial_i g_{jl} + \partial_j g_{il} - \partial_l g_{ij} - ({\Gamma}_{ij}^k+{\Gamma}_{ji}^k) g_{kl} + \delta_{j}^0 F_{il} + \delta_i^0 F_{jl} \right) = 0,  \\
\delta \mathcal{L}_{\rho EE'} = 0 & \implies e_{kEE'} (\Gamma^k_{ji} - \Gamma^k_{ij}) = 0.
\end{align}
These are precisely the equations which we wrote above in \eqref{covg} and \eqref{torsionfree}. We have now shown that they arise naturally by demanding that the 1-fermion terms in the SUSY variation of the deformed sigma model vanish.

\subsection{Supersymmetry for boundaries}\label{boundarySUSY}

Now suppose we consider the deformed \hk sigma model on a space with boundary,
such as we will encounter upon integrating out a NUT center as described
in the introduction.  The deformed action including the boundary terms
should still have the full supersymmetry which we have before integrating out.
In this section we work out the condition this imposes on the boundary terms.

Let us consider what the boundary term in the action should look like if we restrict to constant fields
(or equivalently consider just the lowest term in the derivative expansion).
Then we will have
\begin{equation}
\mathcal{L}_\partial = H(\varphi)
\end{equation}
for some function $H$ on $\ctM$.  The supersymmetry variation of this term is
\begin{equation}
\delta \mathcal{L}_\partial = \partial_i H \delta_\zeta \varphi^i
\end{equation}
which expands out to
\begin{equation}
\delta \mathcal{L}_\partial = \partial_i H (\psi^{E'\alpha} \bar\zeta^E_\alpha e^i_{EE'} + \bar\psi_{E'\alpha} \zeta^\alpha_E e^{i E E'})
\end{equation}
This has to be added to extra boundary terms coming from integration by parts in the bulk variation.
These arise only in the 1-fermion terms; the terms involving $\bar\zeta_E$ are
\begin{subequations}
\begin{align}
4 \pi \delta \mathcal{L}_{ibp} =\;
&\psi^{E'\alpha} \bar\zeta^E_\alpha e^i_{EE'} \partial_N \varphi^j g_{ij} \\
-& \partial_\nu \varphi^i e_{iEA'} \gamma_{\sigma\alpha}^{\nu} \bar\zeta^{\sigma E} \gamma^{\alpha}_{N\beta} \psi^{A'\beta} \\
+& \half \eps_{\mu \nu N} G^{\mu\nu} A_i \psi^{E'\alpha} \bar\zeta^E_\alpha e^i_{EE'}
\end{align}
\end{subequations}
where here and below, $N$ denotes a unit normal to the boundary.
Collecting all terms involving $\bar\zeta_E$ we have
\begin{subequations}
\begin{align}
4 \pi \delta \mathcal{L} =\; &4 \pi \delta \mathcal{L}_\partial + 4 \pi \delta \mathcal{L}_{ibp}\\
=\;& 4 \pi \partial_i H \psi^{E'\alpha} \bar\zeta^E_\alpha e^i_{EE'} \\
+& \psi^{E'\alpha} \bar\zeta^E_\alpha e^i_{EE'} \partial_N \varphi^j g_{ij} \\
-& \partial_\nu \varphi^i e_{iEA'} \gamma_{\sigma\alpha}^{\nu} \bar\zeta^{\sigma E} \gamma^{\alpha}_{N\beta} \psi^{A'\beta} \\
+& \half \eps_{\mu \nu N} G^{\mu\nu} A_i \psi^{E'\alpha} \bar\zeta^E_\alpha e^i_{EE'}
\end{align}
\end{subequations}
The above equation can be rewritten as
\begin{subequations}
\begin{align}
&\delta \mathcal{L} = e_{iEE'} \psi^{E' \alpha} \bar\zeta^{E}_\sigma \delta \mathcal{L}_\alpha^{\sigma i}\\
&4 \pi \delta \mathcal{L}_\alpha^{\sigma i} = (4 \pi \partial^i H + \half \eps_{\mu \nu N} G^{\mu\nu} A^i) \delta_\alpha^\sigma + (\partial_N \varphi^i \delta_\alpha^\sigma - \partial_\nu \varphi^i (\gamma^\nu \gamma_N)_{\alpha}^\sigma).
\end{align}
\end{subequations}
Using the condition that the fields are constant over the boundary,
this variation reduces to
\begin{subequations}
\begin{align}
4 \pi \delta \mathcal{L} =
\psi^{E' \alpha} \bar\zeta^{E}_\sigma e_{iEE'} (4 \pi \partial^i H + \half \eps_{\mu \nu N} G^{\mu\nu} A^i) \delta_\alpha^\sigma
\end{align}
\end{subequations}

Evidently $\delta \mathcal{L} =0$ if
\begin{equation}
 e_{iEE'} (\partial^i H + \frac{1}{8 \pi} \eps_{\mu \nu N} G^{\mu \nu} A^i) = 0.
\end{equation}
After integrating over the boundary this condition gives
\begin{equation}
e_{iEE'} (\partial^i Q + k A^i) = 0,
\end{equation}
where $k$ is the degree introduced in \eqref{degree}, and $Q = H \vol(\partial)$.
Finally, let $W = \exp(Q)$; $W$ is the contribution to the path integral integrand
coming from this boundary.  Then $W$ obeys
\begin{equation}
\boxed{e_{iEE'} (\partial^i + k A^i) W = 0} \label{bdrySUSY}
\end{equation}
Recalling that we require this only for $E=2$ (but arbitrary $E'$),
we obtain exactly \eqref{pert-holomorphy} which appeared in the introduction.
In \S\ref{holomorphy-boundary} we will check this equation directly, in
the particular example of a deformed 3D sigma model obtained by compactifying $U(1)$ SYM
on Taub-NUT space.

\section{\tops{$U(1)$}{U(1)} theories compactified on \tops{$S^1$}{S1}}

In this section we consider $U(1)$ super Yang-Mills compactified on $\mathbb{R}^3 \times S^1$ and study the corresponding hyperk\"ahler sigma model as a warm up example before taking on the case of $U(1)$ super Yang-Mills compactified on Gibbons-Hawking space in \S\ref{GH-U(1)ex}. We first discuss the dimensional reduction of a 4-dimensional bosonic $U(1)$ gauge theory and dualization of the resulting 3D action. We then present the analogous computation for $U(1)$ super Yang-Mills and derive the related \hk sigma model.

\subsection{Bosonic \tops{$U(1)$}{U(1)} gauge theory}\label{U(1)flat}

The metric on $\mathbb{R}^3 \times S^1$ is
\begin{equation}
ds^2_{4}= \sum_i (\de x^i)^2 +R^2 d\chi^2 =\sum_i (\de x^i)^2 +\de y^2, \; y \sim y +2\pi R
\end{equation}

\subsubsection*{Action and dimensionally reduced fields}
The action of a pure $U(1)$ gauge theory on $\mathbb{R}^3 \times S^1$ is given by\footnote{We are denoting the gauge field in the four-dimensional spacetime as $A^{(4)}$, with curvature $F^{(4)}$.  These should not be confused with the
connection over $\ctM$ which appeared in earlier sections, which we denoted as $A$, with curvature $F$.}
\begin{equation}\label{bosact-flat}
\begin{split}
S_{boson} =& \frac{\im \tau}{4\pi}\int_{\mathbb{R}^3 \times S^1} \left[F^{(4)}\wedge \star^{(4)} F^{(4)}\right] + \I \frac{\re \tau}{4\pi}\int_{\mathbb{R}^3 \times S^1} \left[F^{(4)}\wedge F^{(4)}\right] \\
+& \I \frac{\theta_m}{4\pi^2} \int_{S^2_{\infty} \times S^1} \de \chi \wedge F^{(4)}.
\end{split}
\end{equation}
Note that, in addition to the canonical kinetic term for the gauge field and the usual $\theta$
term, we have included an additional boundary term associated with the monopole charge of the gauge field with the coupling $\theta_m$. For any field configuration invariant under translations along $S^1$, the boundary term reduces to $S_b = \I \frac{\theta_m}{2\pi} \int_{S^2_{\infty}} F^{(4)} = \I l \theta_m$ where
$l$ is the monopole number.
$\theta_m$ can therefore be interpreted as a potential conjugate to the monopole number.  In particular,
the exponentiated action is invariant under a shift $\theta_m \to \theta_m + 2 \pi$.

Any 1-form $\alpha^{(4)} \in \Omega^1(\mathbb{R}^3 \times S^1)$ can be expanded as
\begin{equation}
\alpha^{(4)} = \alpha^{(3)}_i \de x^i + \alpha' \de y
\end{equation}
where $\alpha^{(3)} \in \Omega^1(\mathbb{R}^3)$ and $\alpha' \in \Omega^0(\mathbb{R}^3)$.
Expanding the gauge field (1-form) this way, $A^{(4)}= A_i \de x^i + A_y \de y$, one obtains for the curvature 2-form
\begin{equation}
F^{(4)} = (\partial_j A_i - \partial_i A_j) \de x^i \wedge \de x^j + (\partial_i A_y - \partial_y A_i) \de x^i \wedge \de y.
\end{equation}
To dimensionally reduce the theory on $S^1$, we impose the condition that the Lie derivative of the fields
along the Killing vector field $\partial/\partial \chi$ vanishes, i.e.
\begin{equation}
\mathcal{L}_{\partial/\partial \chi} A^{(4)} = 0 \Leftrightarrow \partial_{y} A^{(4)} = 0,
\end{equation}
which implies that the curvature 2-form can be written as
\begin{equation}
F^{(4)} = (\partial_j A_i - \partial_i A_j) \de x^i \wedge \de x^j + \partial_i A_y \de x^i \wedge \de y \equiv F^{(3)} - \de\sigma \wedge \de y,
\end{equation}
where $F^{(3)} \in \Omega^2(\mathbb{R}^3)$ and $\sigma \in \Omega^0(\mathbb{R}^3)$.  $F^{(3)}$ obeys the usual Bianchi identity for a gauge field on $\mathbb{R}^3$, $\de F^{(3)}=0$.

Invariance under large gauge transformations requires that the scalar $\sigma$ be periodic:
\begin{equation}
\sigma \sim \sigma + \frac{1}{2R}.
\end{equation}

\subsubsection*{Star operators}

For any 1-form $\alpha^{(4)}$ on the manifold $\mathbb{R}^3 \times S^1$, one has
\begin{equation}
 \star^{(4)} \alpha^{(4)} = \star^{(3)} \alpha^{(3)} \wedge \de y - \star^{(3)} \alpha'
\end{equation}
where $\star^{(4)}$ and $\star^{(3)}$ are the Hodge star operators on $\R^4$ and $\mathbb{R}^3$
respectively.
Similarly, for any 2-form $\beta^{(4)}$ on $\mathbb{R}^3 \times S^1$ which admits a decomposition as $\beta^{(4)} = \beta^{(3)} + \beta' \wedge \de y$ (which is the case for the curvature 2-form with the condition of dimensional reduction), we have
\begin{equation}
 \star^{(4)} \beta^{(4)} = \star^{(3)} \beta^{(3)} \wedge \de y + \star^{(3)} \beta'.
\end{equation}

\subsubsection*{Dimensionally reduced action and dualization}

Using the above decompositions of 2-forms and star operators, we have
\begin{equation}
\begin{split}
 &\frac{1}{4 \pi} \int_{\mathbb{R}^3 \times S^1} F^{(4)} \wedge \star^{(4)} F^{(4)} = \frac{R}{2}(F^{(3)} \wedge \star^{(3)} F^{(3)} +  \de \sigma \wedge \star^{(3)} \de \sigma), \\
& \frac{1}{4\pi}\int_{\mathbb{R}^3 \times S^1} \left[F^{(4)}\wedge F^{(4)}\right] =-R F^{(3)} \wedge \de \sigma, \\
& \frac{\theta_m}{4\pi^2} \int_{S^2_{\infty} \times S^1}  \de \chi \wedge F^{(4)} = \frac{\theta_m}{2\pi} \int_{S^2_{\infty}} F^{(3)}.
\end{split}
\end{equation}
The 3D action is therefore
\begin{equation}
S^{3D}_{boson}= R\frac{\im\tau}{2}\int_{\mathbb{R}^3} \left[F^{(3)} \wedge \star^{(3)} F^{(3)} +  \de \sigma \wedge \star^{(3)} \de \sigma\right] - \I R {\re\tau}\int_{\mathbb{R}^3} \left[F^{(3)} \wedge \de \sigma \right].
\end{equation}
To introduce a scalar dual to the 3D gauge field, one adds to the Lagrangian a new term
\begin{equation}
{S}^{(3)}_{Lagrange} = -2 \I R \int_{\mathbb{R}^3} \gamma \wedge \de F^{(3)} = -2 \I R \int_{S^2_{\infty}} \gamma F^{(3)} + 2 \I R \int_{\mathbb{R}^3} \de \gamma \wedge F^{(3)} . \label{Lagmult}
\end{equation}
Note that the equation of motion for $\gamma$ imposes the Bianchi identity $\de F^{(3)} = 0$.

One needs to make sure that periods of $F^{(3)}$ and $\de \gamma$ over respective $p$-cycles are adjusted such that the Lagrange multiplier term in \eqref{Lagmult} does not change the path integral. If we require that the $\gamma$ is periodic,
\begin{equation}
\gamma \sim \gamma + \frac{1}{2 R},
\end{equation}
then the periods of $F^{(3)}$ over 2-cycles must be integer multiples of $2\pi$, as desired.

One can then integrate out $F^{(3)}$ (considered as an arbitrary $2$-form) using its equation of motion,
\begin{equation}
F^{(3)} = - \I(\im \tau)^{-1} \star^{(3)}(\de \gamma - (\re \tau)\de \sigma),
\end{equation}
to obtain the dualized action:
\begin{equation}
\begin{split}
S^{(3D)}_{dual} &=  \frac{R}{2} \int_{\mathbb{R}^3}\frac{1}{\im \tau} \abs{\de \gamma - \tau \de \sigma}^2 + S^{(3D)}_{{bdry}}, \\
S^{(3D)}_{{bdry}}&= \I \frac{\theta_m}{2\pi} \int_{S^2_{\infty}} F^{(3)} - 2 \I R \int_{S^2_{\infty}} \gamma F^{(3)}.
\end{split}
\end{equation}
The boundary term $S^{(3D)}_{{bdry}}$
can be made to vanish by appropriately choosing the boundary configuration of the scalar field $\gamma$. If $\theta_e$ denotes the holonomy of the gauge field along the boundary $S^1$, the boundary conditions on the scalars $\gamma$ and $\sigma$ may be summarized as
\begin{equation}
\begin{split}
\gamma \to \frac{\theta_m}{4\pi R}, \qquad \sigma \to \frac{\theta_e + 2n\pi}{4\pi R},
\end{split}
\end{equation}
where $n$ is an integer.

\subsection{\tops{$\mathcal{N}=2$}{N=2} \tops{$U(1)$}{U(1)} gauge theory}
In this section, we consider a $\mathcal{N}=2$ $U(1)$ SYM on $\mathbb{R}^3 \times S^1$, which we dimensionally reduce along the circle direction to obtain a three dimensional theory. We then dualize the gauge field in exactly the same as shown above to obtain the corresponding \hk sigma model in 3D.

\subsubsection{Dimensional reduction from 6D \tops{$\mathcal{N}=1$}{N=1} super Yang-Mills}

$\mathcal{N}=2$ $U(1)$ SYM on $\mathbb{R}^4$ can be obtained from a $\mathcal{N}=1$ theory on $\mathbb{R}^3 \times S^1 \times T^{1,1}$ using dimensional reduction.  $(x_0,x_5)$ denote coordinates along $T^{1,1}$ while $(x_1,x_2,x_3,x_4)$ are coordinates along $\mathbb{R}^3 \times S^1$.  The Lagrangian of the 6D theory is
\begin{equation}
 \mathcal{L}_6= \frac{1}{g^2_6} \left[\frac{1}{2} F_{MN} F^{MN} + \bar{\psi}_a \Gamma^M \partial_M \psi^a+\frac{1}{2} D_{ab}D^{ab}\right] \label{6DSYMflat-copy}
\end{equation}
where we choose the metric $\eta_{MN}=\left(-1,1,1,1,1,1\right)$.
The fermionic field $\psi_a$ is a symplectic Majorana-Weyl spinor which transforms as a doublet of the
$SU(2)$ $R$-symmetry.

The action in \eqref{6DSYMflat-copy} is invariant under the following SUSY transformation rules:
\begin{equation}
\begin{split}
&\delta A_M=\frac{1}{2}(\bar{\zeta}_a \Gamma_M \psi^a-\bar{\psi}_a \Gamma_M \zeta^a)= -\bar{\psi}_a \Gamma_M \zeta^a \;(\mbox{since} \; \;\bar{\zeta}_a \Gamma_M \psi^a=-\bar{\psi}_a \Gamma_M \zeta^a)\\
 &\delta \psi_a=-\frac{1}{2} F_{MN} \Gamma^{MN} \zeta_a - D_{ab} \zeta^b, \;  \delta \bar{\psi}_a= \frac{1}{2} \bar{\zeta}_a \Gamma^{MN} F_{MN} - D_{ab} \bar{\zeta}^b\\
 &\delta D_{ab}= \bar{\zeta}_a \Gamma^M \partial_M \psi_b + \bar{\zeta}_b \Gamma^M \partial_M \psi_a
\end{split}\label{6Dsusyflat}
\end{equation}
The SUSY parameter $\zeta_a$ is a Grassman-odd symplectic Majorana-Weyl spinor obeying
the Killing spinor equation on $\mathbb{R}^3 \times S^1\times T^{1,1}$ :
\begin{equation}
\partial_M \zeta^a=0 \implies \zeta^a = \mbox{constant}
\end{equation}
Further details on our conventions for 6D spinors and SUSY transformations of 6D $\mathcal{N}=1$ SYM
can be found in Appendix \ref{6Dspinor}.

\subsubsection{4D action and SUSY transformation}

Next we dimensionally reduce the action, demanding that the Lie derivatives
(which in components will just be represented as ordinary derivatives)
of the gauge and fermionic fields along the torus directions vanish. The rules/notation for dimensional reduction from 6D to 4D spinors are given in Appendix \ref{4D/3Dspinors}.

Dimensional reduction gives the action for $\mathcal{N}=2$ SYM in 4D, to which we add the standard theta term as well as a boundary term with coefficient $\theta_m$, just
as we did in the case of the purely bosonic theory.
\begin{equation}
\begin{split}
S^b_{(4)}&=\frac{\im \tau}{2\pi} \int_{\mathbb{R}^3 \times S^1} \de^4x \left[\frac{1}{2} F_{mn}F^{mn} + \partial^m A_5 \partial_m A_5- \partial^m A_0 \partial_m A_0+\frac{1}{2}D_{ab} D^{ab}\right]\\
&+\I \frac{\re \tau}{2\pi}\int_{\mathbb{R}^3 \times S^1} \left[F^{(4)}\wedge F^{(4)}\right] + \I \frac{\theta_m}{4\pi^2} \int_{S^2_{\infty} \times S^1}  \de \chi \wedge F^{(4)}\\
S^f_{(4)}&= \frac{\im \tau}{2\pi} \int_{\mathbb{R}^3 \times S^1} \de^4x\; \left(-2\I\bar{\lambda}_a \bar{\sigma}^m \partial_m \lambda^a\right) \;\;(m,n=1,2,3,4)
\end{split}
\end{equation}

The SUSY transformation rules may be summarized as
\begin{equation}
\begin{split}
& \delta A_m = \I \left(\bar{\zeta}^{\dot{\alpha}}_a \bar{\sigma}_{m,\;\dot{\alpha} \beta}\lambda^{a,\beta}-\zeta_{a,{\alpha}} \sigma^{\alpha \dot{\beta}}_m \bar{\lambda}^a_{\dot{\beta}}\right)\; (m=1,2,3,4)\\
& \delta A_5 = \zeta_{\alpha,a}\lambda^{\alpha,a}-\bar{\zeta}^{\dot{\alpha}}_a\bar{\lambda}_{\dot{\alpha}}^a,\;\delta A_0 =\zeta_{\alpha,a}\lambda^{\alpha,a}+\bar{\zeta}^{\dot{\alpha}}_a\bar{\lambda}_{\dot{\alpha}}^a \\
&\delta \lambda^{a,\alpha} = -\frac{1}{2} F_{mn}(\sigma^{mn})^{\alpha}_{\;\;\beta}\zeta^{\beta, a} - \I\bar{\zeta}^a_{\dot{\alpha}} (\sigma^{m})^{\alpha \dot{\alpha}} \partial_m A_{5}-\I\bar{\zeta}^a_{\dot{\alpha}} (\sigma^{m})^{\alpha \dot{\alpha}} \partial_m A_{0}  -D^a_{\;\;b}\zeta^{b,\alpha}\\
&\delta \bar{\lambda}^a_{\dot{\alpha}} = -\frac{1}{2} F_{mn}(\bar{\sigma}^{mn})_{\dot{\alpha}}^{\;\;\dot{\beta}}\bar{\zeta}^{a}_{\dot{\beta}} - \I{\zeta}^{a,\;{\alpha}} (\bar{\sigma}^{m})_{\dot{\alpha} {\alpha}} \partial_m A_{5}+\I{\zeta}^{a,\;{\alpha}} (\bar{\sigma}^{m})_{\dot{\alpha} {\alpha}} \partial_m A_{0}-D^a_{\;\;b} \bar{\zeta}^b_{\dot{\alpha}}\\
&\delta D_{ab} = \I\bar{\zeta}^{\dot{\alpha}}_a \bar{\sigma}^m_{\dot{\alpha}\alpha} D_m \lambda^{\alpha}_b +\I \bar{\zeta}^{\dot{\alpha}}_b \bar{\sigma}^m_{\dot{\alpha}\alpha} D_m \lambda^{\alpha}_a -\I{\zeta}_{a,{\alpha}}( {\sigma}^{m})^{\alpha\dot{\alpha}} D_m \bar{\lambda}_{b,\dot{\alpha}}-\I{\zeta}_{a,{\alpha}}( {\sigma}^{m})^{\alpha\dot{\alpha}} D_m \bar{\lambda}_{b,\dot{\alpha}}
\end{split} \label{susy4D-4Dspinorsflat}
\end{equation}

Note that both the theta term and our extra boundary term
are separately invariant under SUSY.

\subsubsection{3D action and dualization}

Dimensional reduction along a circle direction is straightforward for the bosonic action. The reduction of the fermionic action and the SUSY transformations from 4D to 3D is described in Appendix \ref{4D/3Dspinors}.  The 3D action, leaving out the auxiliary fields, is
\begin{equation}
\begin{split}
S^b_{(3)} &= {R \im \tau} \int_{\mathbb{R}^3} \de^3x \left[\frac{1}{2} F_{ij}F^{ij} +\partial^i\sigma \partial_i \sigma+ \partial^i A_5 \partial_i A_5- \partial^i A_0 \partial_i A_0\right]\\
&-\I R \re \tau \int_{\mathbb{R}^3} \left[2 F^{(3)} \wedge \de \sigma \right] + \I \frac{\theta_m}{2\pi} \int_{S^2_{\infty}}  F^{(3)},\\
S^f_{(3)} &= -2 \I R \im \tau \int_{\mathbb{R}^3} \de^3x\; \bar{\lambda}^{\alpha}_a ({\gamma}^i)_{\alpha \beta}\partial_i\lambda^{a\;\beta}=2\I R \im \tau \int_{\mathbb{R}^3} \de^3x\; \bar{\lambda}_a {\gamma}^i\partial_i\lambda^a.
\end{split}
\end{equation}
where $i,j=1,2,3$.

The corresponding SUSY transformation is
\begin{equation}
\begin{split}
&\delta A_i= \I \left({\zeta}_a {\gamma}_{i}\bar{\lambda}^{a}-\bar{\zeta}_a {\gamma}_{i}\lambda^{a}\right)\; (i=1,2,3),\\
&\delta \sigma=(\zeta_a \bar{\lambda}^a-\bar{\zeta}_a \lambda^a)\\
& \delta A_5 =-( \zeta_{a}\lambda^{a}+\bar{\zeta}_a\bar{\lambda}^a),\;\delta A_0 =-(\zeta_{a}\lambda^{a}-\bar{\zeta}_a\bar{\lambda}^a) \;\\
&\delta \lambda^{\alpha,a} = -\frac{\I}{2} F_{ij}\epsilon^{ijk}(\gamma_{k})^{\alpha}_{\;\;\beta}\zeta^{\beta, a} +\I \partial_i\sigma (\gamma^i)^{\alpha}_{\beta}\zeta^{\beta,a}-\I (\gamma^{i})^{\alpha\beta} \bar{\zeta}^a_{\beta}\partial_i A_{5}-\I (\gamma^{i})^{\alpha\beta} \bar{\zeta}^a_{\beta}\partial_i A_{0}  \\
& \delta \bar{\lambda}^a_{\alpha} =\frac{\I}{2} F_{ij}\epsilon^{ijk}(\gamma_{k})_{\alpha}^{\;\;\beta}\bar{\zeta}_{\beta}^{ a} +\I \partial_i\sigma (\gamma^i)_{\alpha}^{\beta}\bar
{\zeta}_{\beta}^{a}- \I(\gamma^{i})_{\alpha\beta} {\zeta}^{a,\;\beta}\partial_i A_{5}+ \I (\gamma^{i})_{\alpha\beta} {\zeta}^{a,\;\beta}\partial_i A_{0}
\end{split} \label{susy3D-3Dspinorsflat}
\end{equation}
The bosonic action (along with the $\theta$-term) may be dualized as shown in Section \ref{U(1)flat}. The final form of the 3D action is
\begin{equation}
\begin{split}
S^b_{(3)}=&R \int_{\mathbb{R}^3} \de^3x \left[(\im\tau)^{-1}|\de\gamma-\tau \de\sigma|^2  +(\im\tau) (\partial^i A_5 \partial_i A_5- \partial^i A_0 \partial_i A_0)\right]\\
S^f_{(3)}=& 2\I R\; \im\tau\int_{\mathbb{R}^3} \de^3x\; \bar{\lambda}_a {\gamma}^i\partial_i\lambda^a \;\;(i,j=1,2,3)
\end{split} \label{dual3daction-flat}
\end{equation}
The SUSY transformation rules of the dualized 3D action are
\begin{equation}
\begin{split}
&\delta \gamma= (\tau \zeta_a \bar{\lambda}^a-\bar{\tau}\bar{\zeta}_a \lambda^a),\\
&\delta \sigma=(\zeta_a \bar{\lambda}^a-\bar{\zeta}_a \lambda^a),\\
& \delta A_5 =-( \zeta_{a}\lambda^{a}+\bar{\zeta}_a\bar{\lambda}^a),\;\delta A_0 =-(\zeta_{a}\lambda^{a}-\bar{\zeta}_a\bar{\lambda}^a), \;\\
&\delta \lambda^{\alpha,a} = -(\im \tau)^{-1} (\partial_k\gamma-\tau \partial_k\sigma)(\gamma^{k})^{\alpha}_{\;\;\beta}\zeta^{\beta, a} - \I (\gamma^{i})^{\alpha\beta} \bar{\zeta}^a_{\beta}(\partial_i A_{5}+ \partial_i A_{0}),\\
& \delta \bar{\lambda}^a_{\alpha} =(\im \tau)^{-1} (\partial_k\gamma-\bar{\tau} \partial_k\sigma)(\gamma^{k})_{\alpha}^{\;\;\beta}\bar{\zeta}_{\beta}^{ a}- \I (\gamma^{i})_{\alpha\beta} {\zeta}^{a,\;\beta}(\partial_i A_{5}- \partial_i A_{0}).
\end{split} \label{susy3D-3Dspinors-dual}
\end{equation}
Backgrounds preserving the SUSY, i.e. obeying $\delta (\mbox{fermions})=0$, are
\begin{equation}
\begin{split}
&\partial_k\gamma-\tau \partial_k\sigma =0,\;\;\partial_k\gamma-\bar{\tau} \partial_k\sigma=0 \implies \gamma =\frac{\theta_m}{4\pi R}, \ \sigma = \frac{\theta_e + 2n\pi}{4\pi R}\\
&A_5 + A_0 = \bar{a} \; (\mbox{constant}), \; A_5 - A_0 = a \; (\mbox{constant})
\end{split}
\end{equation}

\subsection{\tops{$U(1)$}{U(1)} SYM on \tops{$\mathbb{R}^3 \times S^1$}{R3 x S1}:  hyperk\"ahler sigma model picture}\label{U1onR3xS1HK}

The dualized 3D action obtained above is an elementary example of a hyperk\"ahler sigma model in 3D. In the context of the deformed \hk sigma model proposed in this paper, this is an example of
the ``undeformed'' case.

To recast the above 3d action into the standard form of a hyperk\"ahler sigma model,
we organize the scalar fields ($\varphi^i$ with $i=1,2,3,4$) as
\begin{equation}
\begin{split}
&y=R(\gamma -\tau \sigma),\; \bar{y}=R(\gamma -\bar{\tau} \sigma),\\
&\bar{\phi}=(A_5+ A_0),\; {\phi}=(A_5-A_0).
\end{split} \label{redefhkflat}
\end{equation}
The SUSY transformations then reduce to
\begin{equation}\label{SUSYdualU(1)}
\begin{split}
&\delta y= R (\tau-\bar{\tau})\bar{\zeta_a}\lambda^a \;,\; \delta \bar{y}=R(\tau-\bar{\tau})\zeta_a \bar{\lambda}^a\\
& \delta {\phi} =-2\bar{\zeta_{a}}\bar{\lambda^{a}},\;\delta \bar{\phi} =-2\zeta_{a}\lambda^{a} \;\\
&\delta \lambda_{\alpha}^{a} = -(R \im \tau)^{-1} \partial_\mu y(\gamma^{\mu})_{\alpha}^{\;\;\beta}\zeta_{\beta}^ {a}-\I(\gamma^{i})_{\alpha}^{\beta} \bar{\zeta}^a_{\beta}\partial_i\bar{\phi}\\
&\delta \bar{\lambda}^a_{\alpha} =(R\im \tau)^{-1} \partial_k\bar{y} (\gamma^{k})_{\alpha}^{\;\;\beta}\bar{\zeta}_{\beta}^{ a}+\I (\gamma^{i})_{\alpha}^{\beta} {\zeta}^{a}_{\beta} \partial_i {\phi}
\end{split}
\end{equation}
From the dualized action \eqref{dual3daction-flat} and the redefinition \eqref{redefhkflat}, the bosonic part of the action can  be written as
 \begin{equation}
\begin{split}
&S_b= \frac{1}{8\pi} \int_{\mathbb{R}^3} \de^3x \left[ g_{ij} \partial_{\mu} \phi^i \partial^{\mu} \phi^j + \epsilon_{\mu\nu\rho} G^{\mu\nu} \partial^{\rho}\phi^i A_i \right]\\
&g_{\phi\bar{\phi}} =-{N}={4 \pi R \im \tau}\\
&g_{y\bar{y}} =-\tilde{N}=\frac{4 \pi}{R \im \tau}\\
&A_i=0\\
&G=0\\
&\Omega_{A'B'C'D'}=0
\end{split}
\end{equation}

We now show how the fermions and the SUSY parameters in the UV Lagrangian of a $U(1)$ SYM are related to the fermions and the SUSY parameters respectively in the corresponding HK sigma model. In particular, the UV Lagrangian has a manifest $SU(2)$ R-symmetry which is not easily visible in the sigma model description. Relating the fermions on the two sides, among other things, clarifies the action of R-symmetry on the sigma model fermions.

The $Sp(r)$ indices (primed indices) are raised and lowered by the following 2-form
\begin{equation}
\begin{split}
\epsilon_{A' B'} = \begin{pmatrix} 0 & N \\ -N & 0 \end{pmatrix},\; \epsilon^{A' B'} = \begin{pmatrix} 0 & -\frac{1}{N} \\ \frac{1}{N} & 0 \end{pmatrix}
\end{split}
\end{equation}
For the $Sp(1)$ indices (unprimed indices), the corresponding 2-forms are
\begin{equation}
\begin{split}
\epsilon_{A B} = \begin{pmatrix} 0 & -1 \\ 1 & 0 \end{pmatrix},\;  \epsilon^{A B} = \begin{pmatrix} 0 & 1 \\ -1 & 0 \end{pmatrix}\\
\end{split}
\end{equation}

The intertwiner can be explicitly written as
\begin{equation}
\begin{split}
& e^{AA'}_i=\begin{pmatrix} d\phi & \I \tilde{N} d\bar{y} \\ -\I\tilde{N} d{y} & -d\bar{\phi} \end{pmatrix}, \;  e_{i\;AA'}=N\begin{pmatrix} d\bar{\phi} & -\I\tilde{N} \de y \\ \I \tilde{N} d\bar{y} & -d{\phi} \end{pmatrix}\\
&e^{AA'}_0=0,\;  e_{0AA'}=0
\end{split}
\end{equation}
Now writing the SUSY variations of the fermionic fields in the HK sigma model, we have
\begin{equation}\label{SUSYfromHK}
\begin{split}
\delta_\zeta \psi^{1'}_\alpha &= -\I \partial_\nu \phi (\gamma^{\nu})_\alpha^{ \sigma} \zeta^2_{\sigma} -\tilde{N}  \partial_\nu \bar{y} (\gamma^{\nu})_\alpha^{\sigma} \zeta^1_{\sigma} , \\
\delta_\zeta \psi^{2'}_\alpha &= -\tilde{N}  \partial_\nu {y} (\gamma^{\nu})_\alpha^{\sigma} \zeta^2_{\sigma}-\I \partial_\nu \bar{\phi} (\gamma^{\nu})_\alpha^{\sigma} \zeta^1_{\sigma} , \\
\delta_\zeta \bar \psi_{\alpha}^ {1'} &=  -\tilde{N}  \partial_\nu \bar{y} (\gamma^{\nu})_\alpha^{\sigma} \bar\zeta^{1}_{\sigma} - \I\partial_\nu {\phi} (\gamma^{\nu})_\alpha^{\sigma} \bar\zeta^{2}_{\sigma},\\
\delta_\zeta \bar \psi_{\alpha}^ {2'} &=  -\I \partial_\nu \bar{\phi} (\gamma^{\nu})_\alpha^{\sigma} \bar\zeta^{1}_{\sigma}- \tilde{N}  \partial_\nu {y} (\gamma^{\nu})_\alpha^{\sigma} \bar\zeta^{2}_{\sigma}.
\end{split}
\end{equation}

The fermions and SUSY parameters can now be easily related by comparing equation \ref{SUSYdualU(1)} and \ref{SUSYfromHK}.
\begin{equation}
\begin{split}
& \lambda^a_{\alpha} =\frac{1}{\sqrt{2}} \left(\psi^{2'}_\alpha,\bar \psi_{\alpha}^ {2'} \right)\\
& \bar{\lambda}^a_{\alpha}=\frac{1}{\sqrt{2}} \left(\psi^{1'}_\alpha, \bar \psi_{\alpha}^ {1'} \right)\\
& \zeta^a_{\alpha}= \frac{1}{\sqrt{2}} \left(-\zeta^2_{\alpha }, -\bar{\zeta}^2_{\alpha}\right)\\
& \bar{\zeta}^a_{\alpha}=\frac{1}{\sqrt{2}} \left(\zeta^1_{\alpha}, \bar{\zeta}^1_{\alpha}\right)
\end{split}
\end{equation}
One can readily check that SUSY variations of the bosons match precisely with the above identification.\\
\begin{equation}
\begin{split}
&\delta \phi =- \psi^{1'\alpha} \bar{\zeta}^1_{\alpha} - \bar \psi_{\alpha}^ {1'} \zeta^{1\alpha} :=-2\bar{\zeta}_a \bar{\lambda}^a\\
&\delta \bar{\phi}=\psi^{2'\alpha} \bar{\zeta}^2_{\alpha} + \bar \psi_{\alpha}^ {2'} \zeta^{2\alpha}:=-2 \zeta_a \lambda^a\\
&\delta y=-\I N \psi^{2'\alpha} \bar{\zeta}^1_{\alpha}+\I N\bar \psi_{\alpha}^ {2'} \zeta^{1\alpha}:=R(\tau -\bar{\tau})\bar{\zeta}_a \lambda^a\\
&\delta \bar{y}=\I N \psi^{1'\alpha} \bar{\zeta}^2_{\alpha}-\I N\bar \psi_{\alpha}^ {1'} \zeta^{2\alpha}:=R(\tau -\bar{\tau}){\zeta}_a \bar{\lambda}^a
\end{split}
\end{equation}
Therefore one can derive that
\begin{equation}
\begin{split}
&-\I \bar\psi_{\alpha A'} \gamma^{\mu \alpha}_\beta (\partial_\mu \psi^{A'\beta} + q^{A'}_{B' i} \partial_\mu \phi^i \psi^{B' \beta})\\
&=2\I R \im \tau  \bar{\lambda}_a {\gamma}^{\mu}\partial_{\mu}\lambda^a   -\I q^{A'}_{B' i}\bar\psi_{A'} \gamma^{\mu} \partial_\mu \phi^i \psi^{B'}
\end{split}
\end{equation}
which shows that $q^{A'}_{B' i} =0, \; \forall i$.

\section{\tops{$U(1)$}{U(1)} theories on Gibbons-Hawking spaces} \label{GH-U(1)ex}
In this section, we present a nontrivial example of the deformed \hk sigma model introduced in \S\ref{sec:deformed-sigma-model}. We start with $U(1)$ SYM on a general  Gibbons-Hawking space and dimensionally reduce along the circle fiber to obtain an explicit form for the deformed \hk sigma model in 3D. This allows one to compute the connection $\Gamma$ on the family of \hk manifolds $\ctM$ and directly check that the condition of supersymmetry derived for a generic sigma model in this class in \S\ref{sec:deformed-sigma-model} holds in this particular case.

\subsection{Bosonic \tops{$U(1)$}{U(1)} gauge theory on Gibbons-Hawking space}

The action of a bosonic $U(1)$ gauge theory on a Gibbons-Hawking space $X$ is
\begin{equation}\label{bosact-GH}
S_{boson}= \frac{\im \tau}{4\pi}\int_{X} \left[F^{(4)}\wedge \star^{(4)} F^{(4)}\right] + \I \frac{\re\tau}{4\pi}\int_{X} \left[F^{(4)}\wedge F^{(4)}\right].
\end{equation}
The metric on the 4-manifold $X$ can be written in the form
\begin{equation}
ds^2_{X}= V\sum_i (\de x^i)^2 +R^2 V^{-1} \Theta^2,
\end{equation}
where $\Theta= \de \chi + B$, with $B \in \Omega^1(\mathbb{R}^3)$ and $\star^{(3)} \de B =\frac{1}{R} \de V$. Since our task is to reduce the action to flat 3D, we express all $p$-forms in terms of the orthogonal basis of $\de x^1,\de x^2,\de x^3$ and $\Theta$ and rewrite four-dimensional star operators in terms of the three-dimensional (flat) ones, as we did before.

For a 1-form $\alpha \in \Omega^1(X)$,
\begin{align}
&\alpha=\alpha^{(3)} + \alpha' \Theta\\
&\star^{(4)}(\alpha)= R\star^{(3)}(\alpha^{(3)})\wedge \Theta -V^2 R^{-1} \star^{(3)} \alpha'
\end{align}
where $\alpha^{(3)} \in \Omega^1(\mathbb{R}^3)$, $\alpha' \in \Omega^0(\mathbb{R}^3)$ and
$\star^{(3)}$ is the star operator for $\mathbb{R}^3$.

Similarly, for a 2-form $\beta \in \Omega^2(X)$, one can show that
\begin{align}
&\beta=\beta^{(3)} + \beta'\wedge \Theta, \\
&\star^{(4)}(\beta) = R V^{-1}\star^{(3)}(\beta^{(3)})\wedge \Theta +V R^{-1} \star^{(3)} \beta',
\end{align}
where $\beta^{(3)}\in \Omega^2(\mathbb{R}^3)$ and $\beta' \in \Omega^1(\mathbb{R}^3)$.

Therefore, writing $F^{(4)}$ as $F^{(4)}= F^{(3)} -R\;\de\sigma \wedge \Theta$, with $\sigma \in \Omega^0(\mathbb{R}^3)$, we have
\begin{align}
&\frac{1}{4\pi}\int_{S^1}\left[F^{(4)}\wedge \star^{(4)} F^{(4)}\right]=R \;V^{-1}F^{(3)}\wedge \star^{(3)} F^{(3)} +R\; \de\sigma \wedge \star^{(3)}\de{\sigma}\\
&\frac{1}{4\pi}\int_{S^1}\left[F^{(4)}\wedge F^{(4)}\right]= -2 R \;F^{(3)} \wedge \de\sigma
\end{align}
Thus the bosonic action \eqref{bosact-GH}, dimensionally reduced to three dimensions, reads
\begin{equation}
S_{boson}=R\;\im \tau \int_{\mathbb{R}^3} \left[V^{-1}F^{(3)}\wedge \star^{(3)} F^{(3)} + \de \sigma \wedge \star^{(3)}\de {\sigma}\right]-\I R\;\re\tau\int_{\mathbb{R}^3} 2F^{(3)}\wedge \de \sigma
\end{equation}
To dualize, we add the term
\begin{equation}
\begin{split}
S_{dual}=&-2 \I R\int_{\mathbb{R}^3}(\gamma dF^{(3)}+R\;\gamma \wedge \de \sigma \wedge \de B)\\
=&2\I R\int_{\mathbb{R}^3}(\de\gamma \wedge F^{(3)}-R\gamma \wedge \de \sigma \wedge dB)-2\I R \int_{\infty} \gamma F^{(3)}.
\end{split}
\end{equation}
The equation of motion for $F^{(3)}$ modulo the boundary term is
\begin{equation}
F^{(3)}=-\I V (\im \tau)^{-1} \star(\de\gamma - (\re \tau) \de\sigma).
\end{equation}
Integrating out $F^{(3)}$ using the above equation of motion, we arrive at the dualized 3D action:
\begin{equation}
S_{boson}= R \int_{\mathbb{R}^3} \left[V (\im \tau)^{-1} |\de\gamma - \tau \de\sigma|^2 \right] -2 \I R^2\int_{\mathbb{R}^3} \gamma \wedge \de\sigma \wedge dB + S_{boundary},
\end{equation}
where
\begin{equation}
S_{boundary} = -2 \I R\int_{\infty} \gamma F^{(3)}.
\end{equation}

\subsubsection*{Adding the $\theta_m$ term}

Consider adding to the 4D action the following boundary term, generalizing \eqref{bosact-flat} above:
\begin{equation}
\Delta S_E = \I \frac{\theta_m}{8\pi^2 R} \int_{{X}_{\infty}} R \Theta \wedge F^{(4)} \label{bdry-NUT}
\end{equation}
where ${X}_{\infty}$ denotes the $S^1$ bundle on $S^2$ at the boundary of the Taub-NUT space, as $r \to \infty$.
Since $F^{(4)}= F^{(3)} + R \, \de\sigma \wedge \Theta$, we have
\begin{equation}
\Delta S_E= \I \frac{\theta_m}{8\pi^2 R} \int_{X_{\infty}} R\Theta \wedge F^{(3)} = \I \frac{\theta_m}{2\pi}  \int_{{\infty}} F^{(3)}.
\end{equation}
$\Delta S_E$ and $S_{boundary}$ will cancel each other if at $r \to \infty$, $\gamma$ and $\sigma$ have the boundary conditions
\begin{equation}
\boxed{\gamma \to \frac{\theta_m}{4\pi R}, \qquad \sigma \to \frac{\left(\theta_e+2n\pi\right)}{4\pi R}}
\end{equation}

Therefore, the final form of the dualized 3D action is
\begin{equation}
\boxed{S_{boson}= R \int_{\mathbb{R}^3} \left[V (\im \tau)^{-1} \lvert \de\gamma - \tau \de\sigma \rvert^2 \right] - 2 \I R^2\int_{\mathbb{R}^3} \gamma \wedge \de\sigma \wedge \de B }.
\end{equation}

Note that the periodicity of $\theta_m$ is a bit subtler than it was in the case of $\R^3 \times S^1$.
In that case we had simply $S_E (\theta_m + 2 \pi) = S_E(\theta_m)$.
In the present case we have instead
\begin{equation} \label{periodicity-gh}
S_E (\theta_m + 2\pi) = S_E (\theta_m) + k \theta_e + 2 \pi n
\end{equation}
where $k$ measures the degree as defined in \eqref{degree}.
To see this, choose a section $s$ of the $S^1$ bundle over the complement of one point in $S^2$;
so $s$ is a 2-manifold sitting inside the boundary ${X}_\infty$.
The boundary $\partial s$ winds $k$ times around one fiber of the circle bundle.
On the other hand, since $s$ has the topology of $\R^2$ we can choose a global potential $A^{(4)}$
along $s$, and thus we get
\begin{equation}
 \int_{{X}^3_{\infty}} \Theta \wedge F^{(4)} = 4 \pi \int_s F^{(4)} = 4 \pi \int_{\partial s} A^{(4)}
= 4 \pi k (\theta_e + 2 \pi m)
\end{equation}
which gives \eqref{periodicity-gh}.

\subsection{\tops{$\mathcal{N}=2$}{N=2}, \tops{$U(1)$}{U(1)} gauge theory on Gibbons-Hawking space}

\subsubsection{Dimensional reduction from 6D \tops{$\mathcal{N}=1$}{N=1} SYM}

$\mathcal{N}=2$ $U(1)$ SYM on $X$ can be obtained from a $\mathcal{N}=1$ theory on $X \times T^{1,1}$ using dimensional reduction.  Let $(x_0,x_5)$ denote coordinates along $T^{1,1}$ while $(x_1,x_2,x_3,x_4)$ are coordinates along $X$. The action of the 6D theory in terms of vierbeins is
\begin{equation}
 \mathcal{S}_6= \frac{1}{g^2_6}\int \mbox{det} [e]\; d^6 x \left[\frac{1}{2} F_{MN} F^{MN} + \bar{\psi}_a \Gamma^M \partial_M \psi^a+\frac{1}{2} D_{ab}D^{ab}\right] \label{6DSYMNUT}
\end{equation}
The above action is invariant under the SUSY transformation
\begin{equation}
\begin{split}
&\delta A_M=\frac{1}{2}(\bar{\zeta}_a \Gamma_M \psi^a-\bar{\psi}_a \Gamma_M \zeta^a)= -\bar{\psi}_a \Gamma_M \zeta^a \;(\mbox{since} \; \;\bar{\zeta}_a \Gamma_M \psi^a=-\bar{\psi}_a \Gamma_M \zeta^a)\\
 &\delta \psi_a=-\frac{1}{2} F_{MN} \Gamma^{MN} \zeta_a - D_{ab} \zeta^b, \;  \delta \bar{\psi}_a= \frac{1}{2} \bar{\zeta}_a \Gamma^{MN} F_{MN} - D_{ab} \bar{\zeta}^b\\
 &\delta D_{ab}= \bar{\zeta}_a \Gamma^M D_M \psi_b + \bar{\zeta}_b \Gamma^M D_M \psi_a
\end{split}\label{6DsusyNUT}
\end{equation}
The SUSY parameter $\zeta_a$ is a Grassman-odd symplectic Majorana-Weyl spinor and is a solution of the Killing spinor equation on $X \times T^{1,1}$, namely $D_M \zeta^a=0$. For the special case of a single-centered Taub-NUT space, we work out the following solution for the Killing Spinor equation in Appendix \S\ref{KSonNUT}:
\begin{empheq}[box=\fbox]{align}
\begin{split}
&\zeta^a=\zeta^a_0\;(\mbox{constant}) ,\\
&\Gamma_1 \Gamma_2\Gamma_3 \Gamma_4 \zeta^a_0=\zeta^a_0.
\end{split}
\end{empheq}
The above equation ensures that exactly half of the original SUSY on flat space is preserved. From a 4D standpoint, the fermionic parameters generating the preserved SUSY on NUT space are constant chiral spinors.

This is indeed the preserved supersymmetry for any Gibbons-Hawking space $X$. The dualized 3D theory on $\mathbb{R}^3$, which we shall discuss momentarily, is the best place to demonstrate this.

\subsubsection{4D action, SUSY and localization equations}

The standard 4D SYM action on $X$ can be obtained by dimensional reduction of the 6D action discussed above. As in the case for $\mathbb{R}^3 \times S^1$, we add the standard topological term and a boundary term to the bosonic action.
\begin{equation}
\begin{split}
S^b_{(4)}=&\frac{\im \tau}{4\pi} \int_{X} \mbox{det}[e] \; \de^4x \left[\frac{1}{2} F_{mn}F^{mn} + \partial^m A_5 \partial_m A_5- \partial^m A_0 \partial_m A_0+\frac{1}{2}D_{ab} D^{ab}\right]\\
&+\I \frac{\im \tau}{4\pi} \int_{X} \mbox{det}[e] \; \de^4x \;F_{mn} \tilde{F}^{mn}+\I \frac{\theta_m}{8\pi^2 R} \int_{S^3_{\infty}} \Theta \wedge F^{(4)}\\
S^f_{(4)}=& \frac{\im \tau}{4\pi} \int_{X} \mbox{det}[e] \;\de^4x\; \left(-2\I \bar{\lambda}_a \bar{\sigma}^m D_m \lambda^a\right) \;\;(m,n=1,2,3,4)
\end{split}
\end{equation}

The SUSY transformation, generated by a chiral half of the supersymmetry parameters on $\mathbb{R}^3 \times S^1$, may be summarized as:
\begin{equation}
\begin{split}
&\delta A_m= -\I\zeta_{a,{\alpha}} \sigma^{\alpha \dot{\beta}}_m \bar{\lambda}^a_{\dot{\beta}}\; (m=1,2,3,4)\\
& \delta A_5 = \zeta_{\alpha,a}\lambda^{\alpha,a},\;\delta A_0 =\zeta_{\alpha,a}\lambda^{\alpha,a} \;\\
&\delta \lambda^{a,\alpha} = -\frac{1}{2} F_{mn}(\sigma^{mn})^{\alpha}_{\;\;\beta}\zeta^{\beta, a} -D^a_{\;\;b}\zeta^{b,\alpha}\\
& \delta \bar{\lambda}^a_{\dot{\alpha}} = - \I{\zeta}^{a,\;{\alpha}} (\bar{\sigma}^{m})_{\dot{\alpha} {\alpha}} \partial_m A_{5}+\I{\zeta}^{a,\;{\alpha}} (\bar{\sigma}^{m})_{\dot{\alpha} {\alpha}} \partial_m A_{0}\\
& \delta D_{ab}=-\I{\zeta}_{a,{\alpha}}( {\sigma}^{m})^{\alpha\dot{\alpha}} D_m \bar{\lambda}_{b,\dot{\alpha}}-\I{\zeta}_{a,{\alpha}}( {\sigma}^{m})^{\alpha\dot{\alpha}} D_m \bar{\lambda}_{b,\dot{\alpha}}
\end{split} \label{susy4D-4DspinorsNUT}
\end{equation}
To ensure the convergence of the 4D path integral, one needs to consider the theory reduced from a Euclidean version of the 6D theory and this can be achieved by setting $A_0 =\I A^E_0$, with $A^E_0$ real.

With this modification, the localization equations give the following solutions for the bosonic fields:
\begin{equation}
\begin{split}
\boxed{\begin{gathered}F=-\star^{(4)} F\\ A_5 + \I A^E_0 = \bar{a}\;(\mbox{constant}), \\A_5 - \I A^E_0 = a\; (\mbox{constant}) \end{gathered}}
\end{split} \label{loceqn4d}
\end{equation}

\subsubsection{3D action: 4D instanton and Bogomolny equations}

The 3D action may be obtained by dimensional reduction of an $\mathcal{N}=1$ theory on $X \times T^{1,1}$ as shown in the previous section, giving
\begin{equation}
\begin{split}
S^b_{(3)}=&R \;\im\tau \int_{\mathbb{R}^3} \de^3x \left[\frac{1}{2}V^{-1} F^{(3)}_{ij}F^{(3)\;ij} +\partial^i\sigma \partial_i \sigma+ \partial^i A_5 \partial_i A_5- \partial^i A_0 \partial_i A_0\right]\\
&+ \I R\;\frac{\re \tau}{2}   \int_{\mathbb{R}^3} \de^3x \epsilon^{ijk} F^{(3)}_{ij} \partial_k\sigma +\I \frac{\theta_m}{2\pi}\int_{S^2_{\infty}}F^{(3)} \\
S^f_{(3)}=& 2\I R\; \im \tau\int_{\mathbb{R}^3} \de^3x\;  \sqrt{V}\bar{\lambda}^{\alpha}_a ({\gamma}^i)_{\alpha}^{\;\;\beta}\partial_i\lambda^a_{\beta}=2\I R\; \im \tau\int_{\mathbb{R}^3} \de^3x\; \sqrt{V}\bar{\lambda}_a {\gamma}^i\partial_i\lambda^a
\end{split}
\end{equation}
Note that $F^{(3)}$ is not the curvature of a gauge field in three dimensions, since
\begin{equation}
F^{(3)} = \de A-R\sigma \de B =: F - R\sigma dB.
\end{equation}

The SUSY transformations can be summarized as
\begin{equation}
\begin{split}
\delta A_i=& \I\sqrt{V}\zeta_{a} \gamma_i \bar{\lambda}^a\; (i=1,2,3),\; \delta \sigma=\frac{1}{\sqrt{V}}\zeta_a \bar{\lambda}^a\\
 \delta A_5 =&-\zeta_{a}\lambda^{a},\;\delta A_0 =-\zeta_{a}\lambda^{a} \;\\
\delta \lambda^{\alpha,a} =& -\frac{\I}{2} V^{-1} \;F^{(3)}_{ij}\epsilon^{ijk}(\gamma_{k})^{\alpha}_{\;\;\beta}\zeta^{\beta, a} +\I \partial_i\sigma (\gamma^i)^{\alpha}_{\beta}\zeta^{\beta,a}  \\
=& \left(-\frac{\I}{2} V^{-1} \;F_{ij}\epsilon^{ijk}+\I V^{-1}\sigma \partial_k V +\I \partial_k \sigma\right) (\gamma^k)^{\alpha}_{\beta}\zeta^{\beta,a}  \\
\delta \bar{\lambda}^a_{\alpha} =&- \frac{\I}{\sqrt{V}}(\gamma^{i})_{\alpha\beta} {\zeta}^{a,\;\beta}\partial_i( A_{5}- A_{0} )
\end{split} \label{susy3D-3DspinorsNUT}
\end{equation}
The condition $\delta \lambda^{\alpha,a}=0$ gives a modified version of the Bogomolny equation:
\begin{equation}
\frac{1}{2} V^{-1} \;\epsilon^{ijk}F_{ij}=\left(\partial_k \sigma + \sigma \partial_k \log{V}\right) \implies V^{-1}\star_{(3)} F=\de \sigma +\sigma \wedge \de \log V
\end{equation}
Therefore, the localization equations lead to the following solution for the bosonic fields
\begin{equation}
\begin{split}
\boxed{\begin{gathered}V^{-1}\star^{(3)} F=\de \sigma +\sigma \wedge \de \log V\\ A_5+\I A^E_0=\bar{a}, \; A_5-\I A^E_0={a} \end{gathered}}
\end{split} \label{loceqn3d}
\end{equation}
after making the substitution $A_0 \to \I A^E_0$ to ensure the convergence of the path integral, as mentioned earlier.

Note that the equations \ref{loceqn4d} and \ref{loceqn3d} are consistent. Recalling the decomposition of the four-dimensional star operator in terms of the three-dimensional star operator, we have
\begin{equation}
\begin{split}
&\star^{(4)} F^{(4)}= R V^{-1} \star^{(3)} \left(F - R\sigma \de B\right) \wedge \Theta - V \star^{(3)} \de \sigma=-F^{(4)}\\
&\implies \frac{1}{V}  \star^{(3)} \left(F - R\sigma \de B\right) =-\de \sigma \\
&\implies V^{-1}\star^{(3)} F=\de \sigma +\sigma \wedge \de \log V
\end{split}
\end{equation}
which shows that the modified Bogomolny equation obtained in 3D is equivalent to the 4D instanton (solution of the anti-self dual equation) on $X$.

\subsubsection*{Dualized 3D action and SUSY}

The bosonic part of the action may be dualized as before.
\begin{equation}
\begin{split}
S^b_{(3)} &= R \;\int_{\mathbb{R}^3} \left[V (\im \tau)^{-1} |\de\gamma - \tau \de\sigma|^2+ \im \tau(\de A_5 \wedge \star \de A_5 +  \de A^E_0 \wedge \star \de A^E_0)\right]\\
&\;\;\;  -2\I R^2\int_{\mathbb{R}^3} \gamma \wedge \de\sigma \wedge dB.\\
S^f_{(3)} &= 2\I R\; \im \tau\int_{\mathbb{R}^3} \de^3x\;  \sqrt{V}\bar{\lambda}^{\alpha}_a ({\gamma}^i)_{\alpha}^{\;\;\beta}\partial_i\lambda^a_{\beta}=2\I R\; \im \tau\int_{\mathbb{R}^3} \de^3x\; \sqrt{V}\bar{\lambda}_a {\gamma}^i\partial_i\lambda^a. 
\end{split}\label{GH3d-dual}
\end{equation}
where $i,j=1,2,3$ and the rules for SUSY are as follows:
\begin{equation}
\begin{split}
&\delta \gamma=\frac{1}{\sqrt{V}}\tau\zeta_a \bar{\lambda}^a \;,\; \delta \sigma=\frac{1}{\sqrt{V}}\zeta_a \bar{\lambda}^a\\
 &\delta A_5 =-\zeta_{a}\lambda^{a},\;\delta (\I A^E_0) =-\zeta_{a}\lambda^{a} \;\\
&\delta \lambda_{\alpha}^{a} = -(\im \tau)^{-1} \partial_k(\gamma - \tau \;\sigma)(\gamma^{k})^{\alpha}_{\;\;\beta}\zeta^{\beta, a}\\
&\delta \bar{\lambda}^a_{\alpha} =-\frac{\I}{\sqrt{V}}(\gamma^{i})_{\alpha}^{\;\;\beta} {\zeta}^{a}_{\beta}\partial_i( A_{5}- \I A^E_{0} )
\end{split} \label{susydual3D-3DspinorsGH}
\end{equation}
Note that the 3D dualized action in \eqref{GH3d-dual} follows from dimensional reduction of  $U(1)$ SYM on a generic Gibbons-Hawking space parametrized by the scalar function $V$ and the 1-form $B$ (and not just NUT space). One can then directly check that this action is invariant under SUSY rules summarized in \eqref{susydual3D-3DspinorsGH} for a constant $\zeta_a$. Therefore a general Gibbons-Hawking space preserves exactly the same supersymmetry as a NUT space (Appendix \ref{KSonNUT}).

The localization equations for the dualized 3D action can be read off from \eqref{susydual3D-3DspinorsGH}.
\begin{equation} \label{loceq}
\begin{split}
&\de \gamma - \tau \;\de \sigma = 0, \\ & F-R\sigma \de B = - \I V (\im\tau)^{-1} \star^{(3)}(\de\gamma -(\re \tau) \de\sigma), \\& A_5 + \I A^E_0=\bar{a},\; A_5 - \I A^E_0={a}.
\end{split}
\end{equation}
The equations \eqref{loceq} lead to the following first order differential equations for $\gamma$ and $\sigma$:
\begin{equation}
\begin{split}
\boxed{\begin{gathered}V^{-1}\star^{(3)}F=\de \sigma +\sigma \wedge \de \log{V}, \\ \de \gamma - \tau \;\de \sigma =0, \\ \gamma
\to \frac{\theta_m}{4\pi R}, \quad \sigma \to \frac{(\theta_e +2n\pi)}{4\pi R} \; (r\to \infty) \end{gathered}}
\end{split} \label{gammasigma}
\end{equation}

\subsection{\tops{$U(1)$}{U(1)} SYM on Gibbons-Hawking Space: hyperk\"ahler sigma model picture}\label{HKNUT}

The dualized 3D action obtained above is an elementary example of the deformed hyperk\"ahler sigma model introduced in \S\ref{defaction-hk}. To recast the above 3D action into the standard form of a hyperk\"ahler sigma model action,  we organize the scalar fields ($\varphi^i$ with $i=1,2,3,4$)  in the following manner:
\begin{equation}
\begin{split}
&y=R(\gamma -\tau \sigma),\; \bar{y}=R(\gamma -\bar{\tau} \sigma),\\
&\phi=A_5-\I A^E_0,\; \bar{\phi}=A_5+ \I A^E_0.
\end{split}
\end{equation}
SUSY transformations then reduce to the following form
\begin{equation}
\begin{split}
&\delta y=0 \;,\; \delta \bar{y}=\frac{R({\tau}-\bar{\tau})}{\sqrt{V}}\zeta_a \bar{\lambda}^a\\
& \delta {\phi} =0,\;\delta \bar{\phi} =-2\zeta_{a}\lambda^{a} \;\\
&\delta \lambda_{\alpha}^{a} = -(R \im \tau)^{-1} \partial_\mu y(\gamma^{\mu})_{\alpha}^{\;\;\beta}\zeta_{\beta}^ {a}\\
&\delta \bar{\lambda}^a_{\alpha} =\frac{\I}{\sqrt{V}}(\gamma^{\mu})_{\alpha}^{\;\;\beta} {\zeta}^{a}_{\beta}\partial_{\mu}{\phi}
\end{split} \label{3Dsusyhk}
\end{equation}
Defining $\varphi^0=\frac{V}{R}$, the bosonic part of the action can therefore be written as
 \begin{equation}
\begin{split}
&S_b= \frac{1}{8 \pi} \int_{\mathbb{R}^3} \de^3x \left[ g_{ij} \partial_{\mu} \phi^i \partial^{\mu} \phi^j +  \epsilon_{\mu\nu\rho} G^{\mu\nu} \partial^{\rho}\phi^i A_i\right]\\
&g_{\phi\bar{\phi}} = -N = 4 \pi R \im \tau,\\
&g_{y\bar{y}} =-\tilde{N}=\frac{4 \pi V}{R \im \tau},\\
&g_{00}=0,\\
&g_{0i}=0,\\
&\Omega_{A'B'C'D'}=0.
\end{split}
\end{equation}

Unlike the case of $U(1)$ SYM on $\mathbb{R}^3 \times S^1$, the connection $A$ on the line bundle $\mathcal{L}$ is nontrivial in this case. The nonzero components of the connection and the curvature are
\begin{empheq}[box=\fbox]{align}
\begin{split}
&A_y= -A_{\bar{y}}=  8 \I \pi \frac{(\tau \bar{y} - \bar{\tau} y)}{(\tau-\bar{\tau})^2}\\
&F_{y\bar{y}}= -\frac{8\I \pi}{(\tau-\bar{\tau})}\\
\end{split}
\end{empheq}

As in the case of $U(1)$ SYM on $\mathbb{R}^3\times S^1$, $Sp(r)$ indices (primed indices) are raised and lowered by the antisymmetric pairing
\begin{equation}
\begin{split}
\epsilon_{A' B'} = \begin{pmatrix} 0 & N \\ -N & 0 \end{pmatrix},\; \epsilon^{A' B'} = \begin{pmatrix} 0 & -\frac{1}{N} \\ \frac{1}{N} & 0 \end{pmatrix}
\end{split}
\end{equation}
The intertwiner $e$ can be explicitly written as
\begin{equation}
\begin{split}
&e^{AA'}_i=\begin{pmatrix} \de\phi & i\tilde{N} d\bar{y} \\ -\I \tilde{N} \de {y} & -\de\bar{\phi} \end{pmatrix}, \quad \; e_{i\;AA'}=N\begin{pmatrix} \de \bar{\phi} & -i\tilde{N} \de y \\ \I\tilde{N} \de\bar{y} & -\de{\phi} \end{pmatrix}\\
&e^{AA'}_0=0,\; e_{0AA'}=0
\end{split} \label{intertwiners1}
\end{equation}
The fermions and the SUSY parameters in the UV Lagrangian of $U(1)$ SYM may be related to the fermions and the SUSY parameters respectively in the corresponding \hk sigma model.
From the discussion in \S\ref{U1onR3xS1HK}, we find that half of the SUSY parameters have to be set to zero, namely
\begin{equation}
\zeta^1_{\sigma}=0, \; \bar{\zeta}^1_{\sigma}=0.
\end{equation}
The fermions and SUSY parameters can now be easily related:
\begin{equation}
\begin{split}
& \lambda^a_{\alpha} =\frac{1}{\sqrt{2V}} \left(\psi^{2'}_\alpha,\bar \psi_{\alpha}^ {2'} \right),\\
& \bar{\lambda}^a_{\alpha}=\frac{1}{\sqrt{2}} \left(\psi^{1'}_\alpha, \bar \psi_{\alpha}^ {1'} \right),\\
& \zeta^a_{\alpha}= \frac{\sqrt{V}}{\sqrt{2}} \left(-\zeta^2_{\alpha}, -\bar{\zeta}^2_{\alpha}\right).
\end{split}
\end{equation}
With the above identification, one can readily check that the SUSY transformation of the scalars and fermions in the sigma model matches \eqref{3Dsusyhk}.
Since  $\zeta^a_{\alpha}$ is a constant spinor, the above identification immediately implies
\begin{equation}
\begin{split}
&\partial_{\mu} \zeta^E + \frac{1}{2} \frac{\partial_{\mu} V}{V}\zeta^E=0,\\
&\partial_{\mu} \bar{\zeta}^E + \frac{1}{2} \frac{\partial_{\mu} V}{V}\bar{\zeta}^E=0,\\
\implies & f(\varphi^0)=\frac{R}{2V}.
\end{split}
\end{equation}
One can also read off $q^{A'}_{B' i}$ directly by comparing the fermionic actions in the two descriptions:
\begin{equation}
\begin{split}
&S_f=2 \I \im\tau R \int_{\mathbb{R}^3} \de^3 x \sqrt{V} \bar{\lambda}_a {\gamma}^i\partial_i\lambda^a=-\int_{\mathbb{R}^3} \de^3 x\; \bar\psi_{\alpha A'} \gamma^{\mu \alpha}_\beta (\partial_\mu \psi^{A'\beta} + q^{A'}_{B' 0} \partial_\mu V \psi^{B' \beta})\\
&q^{A'}_{B' 0}=\frac{R}{V}\begin{pmatrix}  1/2 & 0  \\ 0 & -1/2 \end{pmatrix},\; q^{A'}_{B' i} =0 \; (i \neq 0)
\end{split}
\end{equation}
Therefore, the effective $\tilde{q}^{A'}_{0B'}$ that appears in the extended \hk identity \eqref{eq:e-cc2}
will be given as
\begin{equation}
\begin{split}
\tilde{q}^{A'}_{B' 0}=\frac{R}{V}\begin{pmatrix}  1 & 0  \\ 0 & 0\end{pmatrix}
\end{split}
\end{equation}
Given the explicit forms of $\tilde{q}^{A'}_{B' 0}$ and $f(\varphi^0)$, one obtains the following nontrivial components of the connection in the extended \hk identity \eqref{eq:e-cc2}:
\begin{equation}
\begin{split}
&\Gamma^{\phi}_{0\phi}= 0, \; \Gamma^{\bar{\phi}}_{0\bar{\phi}}= -\frac{R}{V} \\
&\Gamma^{y}_{0y}= 0, \; \Gamma^{\bar{y}}_{0\bar{y}}= \frac{R}{2V}\\
&\Gamma^{\phi}_{\phi 0}= 0, \; \Gamma^{\bar{\phi}}_{\bar{\phi}0}= 0 \\
&\Gamma^{y}_{y0}= 0, \; \Gamma^{\bar{y}}_{\bar{y}0}= 0\\
&\Gamma^{k}_{00}= 0 \; \forall k.
\end{split} \label{conn1}
\end{equation}
Now we can check whether this connection obeys the constraints arising from the vanishing of 1-fermion terms in the SUSY variation of the action, \eqref{covg}-\eqref{torsionfree}, which we derived for a general deformed \hk sigma model which preserves some supersymmetry.  Note that the 3-fermion constraint \eqref{eq:hk-curvature2} is satisfied trivially in this case.

For $U(1)$ SYM on $X$, the non-trivial part of the first constraint, namely for $j=0$ and arbitrary $i$, assumes the particular form
\begin{equation}
\begin{split}
e^l_{EE'} \left( \partial_i g_{0l} + \partial_0 g_{il} - \partial_l g_{i0} - ({\Gamma}_{i0}^k+{\Gamma}_{0i}^k) g_{kl} +  F_{il}  \right) = 0
\end{split}
\end{equation}
From the structure of the intertwiners specified in \eqref{intertwiners1} and the nonzero components of $\Gamma^k_{ij}$ in \eqref{conn1}, it is clear that there are only two nontrivial components one needs to check, namely for $l=\bar{y}, i=y$, and $l=\bar{\phi}, i=\phi$.

In the first case, we have
\begin{equation}
\begin{split}
&\partial_yg_{0\bar{y}}+\partial_0 g_{y\bar{y}}- \partial_{\bar{y}} g_{y0}-(\Gamma^k_{y0}+\Gamma^k_{0y})g_{k\bar{y}}+ F_{y\bar{y}}\\
&=\partial_0 g_{y\bar{y}}+F_{y\bar{y}}\\
&=\frac{4 \pi }{ \im \tau} -\frac{8\I\pi}{(\tau -\bar{\tau})}=0.
\end{split}
\end{equation}

In the second case , the constraint is satisfied trivially
\begin{equation}
\begin{split}
&\partial_\phi g_{0\bar{\phi}} + \partial_0 g_{\phi\bar{\phi}} -  \partial_{\bar{\phi}} g_{\phi0} -(\Gamma^k_{\phi 0}+ \Gamma^k_{0 \phi})g_{k\bar{\phi}} + F_{\phi \bar{\phi}}\\
&=0
\end{split}
\end{equation}
since each of the terms in the equation is individually zero.

The connection derived in \eqref{conn1} for the \hk sigma model which arises from the
circle compactification of $U(1)$ SYM on $X$ space therefore obeys the first SUSY constraint \eqref{covg}.
Finally, the second SUSY constraint \eqref{torsionfree} can be written as
\begin{align}
&e_{k2E'}(\Gamma_{i0}^k  -\Gamma_{0i}^k)=0\\
\implies & \Gamma_{i0}^y  -\Gamma_{0i}^y=0, \; \Gamma_{i0}^{\phi}  -\Gamma_{0i}^{\phi}=0\; \; \forall i
\end{align}
which is trivially satisfied in this case, since all the relevant components of the connection vanish.

In appendix \ref{U1hkrescaled}, we consider the sigma model again after rescaling the adjoint scalar so that the metric looks closer to the one obtained via compactification on $\mathbb{R}^3 \times S^1$, with an effective radius $R_{eff}=R/V$.

\section{The NUT centers}\label{NUToperator}

\subsection{Setup}

So far we have described the local physics of the
3-dimensional sigma model which one obtains by starting with the
pure $U(1)$ $\N=2$ gauge theory in four dimensions and dimensionally reducing on
a Gibbons-Hawking space $GH$.  Now suppose we consider the actual compactified theory
as opposed to the naive dimensional reduction.
On general grounds we would expect that the local physics of this theory
at energy scales $E \ll V/R$ and $E \ll \norm{\de V}$
can be described by the same fields which appear in
the dimensionally reduced theory.  In fact, here we can say more:
since the four-dimensional theory is free (even on the
Gibbons-Hawking space) the IR physics of the true
compactified theory is governed by the \ti{same} Lagrangian we obtain by
dimensional reduction --- there are no quantum corrections.

More precisely, what we have described so far is the physics in the locus where
$V$ is finite, and hence the fiber of $GH$ is a finite-size circle.  In any complete example
where $V \to 1$ at infinity, $V$ must have singularities, as it is a bounded harmonic function.
We assume $GH$ is smooth; then at these singularities we must have the precise coefficient
$V \sim R/r$ (recall that $R$ is
the asymptotic radius of the circle of $GH$, and $r$ is the distance
from the singularity.)
At these points our dimensional reduction procedure breaks down.

How should these singularities be incorporated in the reduced theory?
We adopt a brutal approach:  cut out a neighborhood of each singularity in $GH$,
of radius $L \gg R$,
and then study the compactified theory at energies $E \ll 1/L \ll 1/R$.  In four-dimensional
terms, the resulting spacetime has a boundary with one $S^3$ component for each singularity
we cut out; in the compactified
theory the corresponding boundary components have the topology of $S^2$.
The physics of the compactified theory is described by the same local Lagrangian as before,
plus some new, unknown boundary interaction at the new $S^2$.
At energy $E \ll 1/L$ this interaction will be well approximated by the leading term
in the derivative expansion, namely the $0$-derivative term,
which we may write as $Q(\varphi)$ for some function $Q$ on $\cM$.

To determine $Q(\varphi)$ explicitly we will compute the partition function $\Psi$ of the
$U(1)$ gauge theory on a particular Gibbons-Hawking space, namely Taub-NUT space,
characterized by the harmonic function $V = 1 + R/r$.

One way of doing this computation is to work directly with the UV description of the theory.
We obtain an answer which in principle can depend on various choices involving the
boundary at spatial infinity:  we have a complex parameter $a$ which gives the asymptotic value of the
complex scalar of the theory, an angle $\theta_e$ which gives the asymptotic value of the
holonomy of the $U(1)$ gauge field around the circle fiber, and a parameter $\theta_m$
which is inserted explicitly into the boundary term \eqref{bdry-NUT} in the action.

On the other hand we can also work with the IR description just discussed.
In this version of the story, the parameters $a, \theta_e, \theta_m$
enter on a more equal footing:  they determine a point of the target $\cM$
of the sigma model, which gives a Dirichlet boundary condition for the sigma model fields
at infinity.  Since we are in the limit $E \ll 1/L$ and the sigma model is IR free,
the partition function up to overall constant
will be simply the contribution from constant fields; and since the bulk action vanishes on
constant fields, the answer will come just from the boundary term on the $S^2$ we have cut out
around the NUT center.  Thus we get
\begin{equation}
\Psi = e^Q.
\end{equation}
Comparing this with the UV computation thus determines $Q$.

\subsection{UV computation}

The bosonic part of the dualized 3D action for a Gibbons-Hawking space $X$ is given as
\begin{equation}
\begin{split}
S^{(0)}_{boson} = & \int_{\mathbb{R}^3} R \left[V (\im \tau)^{-1} |\de\gamma - \tau \de\sigma|^2 + (\im \tau)|\de\phi|^2 \right] - 2\I R^2\int_{\mathbb{R}^3} \gamma \wedge \de\sigma \wedge \de B
\end{split}
\end{equation}
where we have defined the scalar fields $\phi = A_5 - \I A^E_0,\,\bar{\phi} = A_5 + \I A^E_0$.

\subsubsection*{Instanton configurations}

As explained in the previous section, the path integral of $U(1)$ super Yang-Mills on $X$ is completely localized on the following set of instantonic configurations:
\begin{equation}
F^{(4)}_{-}= -\frac{(\theta_e + 2n\pi)}{4\pi} \de \left(\frac{\Theta}{V}\right)=: \alpha \; \de \left(\frac{\Theta}{V}\right)
\end{equation}
In terms of 3D fields, the above configuration has
\begin{equation}
F^{(3)}_{-}=\frac{\alpha}{V} \de B,\; \sigma= -\frac{\alpha}{RV}
\end{equation}
Noting that $F^{(3)} = -\I V (\im \tau)^{-1} \star_3(\de\gamma -(\re\tau) \de\sigma)$ and demanding that $\gamma \to \theta_m/4\pi R$ asymptotically (so that the boundary terms vanish as explained earlier), the corresponding solution for $\gamma$ and $\sigma$ is
\begin{equation}
\begin{split}
\boxed{\begin{gathered}\gamma=\frac{\theta_m -\tau \theta_e}{4 \pi R} -\frac{2\pi n \tau}{4 \pi R} +\frac{(\theta_e + 2n\pi)\tau}{4\pi RV}\\
\sigma=\frac{(\theta_e + 2n\pi)}{4\pi RV} \end{gathered}}
\end{split}
\end{equation}
Note that this is a particular case of \eqref{gammasigma} where $F=0$.

Since $\de(\gamma-\tau \sigma)=0$ for this configuration, the only contribution to the action comes from the topological term.
Now, let us evaluate the action in the special case where $X$ is NUT space:
\begin{equation}
\begin{split}
S^{(n)}_{inst} &= -2\I R^2\int_{\mathbb{R}^3} \gamma \wedge \de\sigma \wedge dB =- 2\I R\int_{\mathbb{R}^3} \gamma \wedge \de\sigma \wedge \star_3 dV\\
&= -2\I R \int_{\mathbb{R}^3} \de^3 x \left(\frac{\theta_m -\tau \theta_e}{4 \pi R} -\frac{2\pi n \tau}{4 \pi R} +\frac{(\theta_e + 2n\pi)\tau}{4\pi RV}\right) \partial_i \left(\frac{(\theta_e + 2n\pi)}{4\pi RV} \right) \partial_i V\\
&= 2\I R \left(\frac{\theta_m -\tau \theta_e}{4 \pi R} -\frac{2\pi n \tau}{4 \pi R}\right)\left(\frac{(\theta_e + 2n\pi)}{4\pi R} \right)\int_{\mathbb{R}^3} \de^3 x \left(\frac{\partial_i V}{V}\right)^2 \\
&+2\I R \tau \left(\frac{(\theta_e+2n\pi)}{4\pi R}\right)^2 \int_{\mathbb{R}^3} \de^3 x \frac{1}{V} \left(\frac{\partial_i V}{V}\right)^2\\
&= \frac{\I}{2\pi} \left[(-\tau \theta_e^2/2+\theta_e \theta_m)+ 2\pi n (\theta_m - \tau \theta_e) -2\pi^2 n^2 \tau \right]
\end{split}
\end{equation}
where in the final step we used $\int_{\mathbb{R}^3} \de^3 x \left(\frac{\partial_i V}{V}\right)^2=4\pi R$ and $\int_{\mathbb{R}^3} \de^3 x \frac{1}{V} \left(\frac{\partial_i V}{V}\right)^2=4\pi (R/2)$.
Thus the partition function comes out to
\begin{equation}
\begin{split}
&\Psi\left(\theta_e,\theta_m,\tau \right)= \sum_{n \in \mathbb{Z}} e^{-S^{(n)}_{inst}}=e^{\frac{\I}{2\pi}(\tau \theta_e^2/2-\theta_e \theta_m)}\sum_{n \in \mathbb{Z}} e^{\I \pi n^2 \tau - 2\pi \I n \left(\frac{\theta_m-\tau \theta_e}{2\pi}\right)}\\
&\implies \boxed{\begin{gathered}\Psi\left(\theta_e,\theta_m,\tau \right)=e^{\frac{\I}{2\pi}(\tau \theta_e^2/2-\theta_e \theta_m)}\sum_{n \in \mathbb{Z}} e^{\I \pi n^2 \tau - 2\pi \I n (2y)} = e^{\frac{\I}{2\pi}(\tau \theta_e^2/2-\theta_e \theta_m)} \Theta (\tau,2y )\end{gathered}}
\end{split}
\end{equation}
where we define $y=\frac{\theta_m-\tau \theta_e}{4\pi}, \;\bar{y}=\frac{\theta_m-\bar{\tau} \theta_e}{4\pi} $. Note that it has the expected periodicity properties:
\begin{equation}
\Psi\left(\theta_e+2\pi,\theta_m,\tau \right)=\Psi\left(\theta_e,\theta_m,\tau \right), \qquad \Psi\left(\theta_e,\theta_m+2\pi,\tau \right)=e^{- \I \theta_e} \Psi\left(\theta_e,\theta_m,\tau \right).
\end{equation}

\subsection{Holomorphy of the boundary terms}\label{holomorphy-boundary}

The above formula for the partition function can now be used to explicitly check the equation for boundary supersymmetry \eqref{bdrySUSY}.
Writing $\Psi\left(\theta_e, \theta_m, \tau \right)$ in terms of the coordinates $\phi, \bar{\phi}, y, \bar{y}$ on $\mathcal{M}$, we get
\begin{equation}
\begin{split}
&\Psi\left(y,\bar{y},\tau \right)=e^{8 \I \pi X(y,\bar{y},\tau)} \Theta (\tau, 2y)\\
&X(y,\bar{y},\tau)=\frac{\tau y^2 -2\bar{\tau}y^2-\tau \bar{y}^2 + 2\bar{\tau} y \bar{y}}{2(\bar{\tau}-\tau)^2}
\end{split}
\end{equation}
Recall the formula for the connection $A_i$ derived in \S\ref{HKNUT}:
\begin{equation}
\begin{split}
&A_y = -A_{\bar{y}} = 8 \I \pi \frac{(\tau \bar{y} - \bar{\tau} y)}{(\tau-\bar{\tau})^2}\\
&A_{\phi}=0,\; A_{\bar{\phi}}=0
\end{split}\label{connA-NUT}
\end{equation}
Given the half supersymmetry which is preserved, \eqref{bdrySUSY} will reduce to
\begin{equation}
\begin{split}
&e_{i2E'}\left(\partial^i +  k A^i\right) \Psi=0\\
\implies &(\partial_{\bar{y}}+k A_{\bar{y}})\Psi=0, (\partial_{\bar{\phi}}+k A_{\bar{\phi}})\Psi=0\\
&k=\frac{1}{4\pi}\int_{S^2} G = -1 \;(\mbox{for NUT space})
\end{split}
\end{equation}
The $\bar{\phi}$ component of the equation is trivially satisfied, since $A_{\bar{\phi}}=0$ and $\Psi$ is independent of $\bar{\phi}$. For the $\bar{y}$ component, we have
\begin{equation}
\begin{split}
\partial_{\bar{y}} \Psi\left(y,\bar{y},\tau \right) & = 8 \pi \I \partial_{\bar{y}}X(y,\bar{y},\tau) e^{\frac{\I}{2\pi}X(y,\bar{y},\tau)}\Theta (\tau, 2y)\\
& =  8 \pi \I \frac{(\bar{\tau} y -\tau \bar{y})}{(\tau -\bar{\tau})^2}\Psi\left(y,\bar{y},\tau \right) \\
& = - k A_{\bar{y}} \Psi\left(y,\bar{y},\tau \right)
\end{split}
\end{equation}
where for the final equality we have used the formula \eqref{connA-NUT} for $A$.

\appendix

\section{\tops{$U(1)$}{U(1)} SYM on NUT space as \hk sigma model:  rescaled version}\label{U1hkrescaled}

In this section, we again consider the \hk sigma model obtained from $U(1)$ SYM on NUT Space via circle compactification, but after rescaling the adjoint scalar so that the metric looks closer to the one
obtained via compactification on $\mathbb{R}^3 \times S^1$.

Defining $\varphi^0=1/R_{eff}=\frac{V}{R}$ the bosonic part of the action can now be written as
\begin{equation}
\begin{split}
&S_b=\frac{1}{8\pi} \int_{\mathbb{R}^3} \de^3x \left[g_{ij} \partial_{\mu} \phi^i \partial^{\mu} \phi^j + \epsilon_{\mu\nu\rho} G^{\mu\nu} \partial^{\rho}\phi^i A_i\right]\\
&g_{\phi\bar{\phi}} =-N= \frac{4\pi R \im \tau}{V}\\
&g_{y\bar{y}} =-\tilde{N}=\frac{4 \pi V}{R \im \tau}\\
&g_{00}= \frac{1}{2} \frac{4\pi R^3}{V^3} \phi \bar{\phi} \im \tau\\
&g_{0\phi}=-\frac{1}{2}\frac{4\pi R^2}{V^2} \bar{\phi} \im \tau\\
&g_{0\bar{\phi}}=-\frac{1}{2}\frac{4\pi R^2}{V^2} \phi \im \tau\\
&G=\de B\\
&A_y= -A_{\bar{y}}=  8 \I \pi\frac{(\tau \bar{y} - \bar{\tau} y)}{(\tau-\bar{\tau})^2}\\
&A_{\phi}=0,\; A_{\bar{\phi}}=0
\end{split}
\end{equation}
The intertwiners can again be explicitly written as
\begin{equation}
\begin{split}
& e^{AA'}_i=\begin{pmatrix} \de \phi & \I \tilde{N} \de \bar{y} \\ -\I\tilde{N} \de{y} & -\de\bar{\phi} \end{pmatrix}, \; e_{i\;AA'}=N\begin{pmatrix} \de\bar{\phi} & -\I\tilde{N} \de y \\ \I\tilde{N} \de\bar{y} & -\de{\phi} \end{pmatrix}\\
& e^{AA'}_0=\begin{pmatrix} -\frac{\phi R}{2V}&0 \\ 0 & \frac{\bar{\phi} R}{2V} \end{pmatrix}, \;  e_{0\;AA'}=N\begin{pmatrix}  -\frac{\bar{\phi} N R}{2V} &0 \\ 0 & \frac{\phi N R}{2V} \end{pmatrix}
\end{split} \label{intertwiners}
\end{equation}

To express the fermionic action and the SUSY transformation in terms of the ``effective" radius $1/\varphi^0$, one needs to rescale the fermionic fields and the Killing spinor in the following way:
\begin{equation}
\begin{split}
&\lambda'^{a}_{\alpha}= V^{3/4} \lambda^{a}_{\alpha}\\
&\bar{\lambda}'^{a}_{\alpha}= V^{3/4} \bar\lambda^{a}_{\alpha}\\
&\zeta'^a_{\alpha}= \frac{\zeta^a_{\alpha}}{V^{1/4}}
\end{split}
\end{equation}
The fermionic action and the rules of SUSY variation in terms of the rescaled fields are
\begin{equation}
\begin{split}
&S_f=2\I\; \im \tau\int_{\mathbb{R}^3} \de^3x\; \frac{R}{V}\left(\bar{\lambda}'_a {\gamma}^i\partial_i\lambda'^a-\frac{3}{2} \frac{\partial_{\mu} V}{V}\bar{\lambda}'_a {\gamma}^{\mu}\lambda'^a\right)\\
&\delta y=0 \;,\; \delta \bar{y}=\frac{R({\tau}-\bar{\tau})}{V}\zeta'_a \bar{\lambda}'^a\\
& \delta {\phi} =0,\;\delta \bar{\phi} =-2\zeta'_{a}\lambda'^{a} \\
&\delta {\lambda}'^{a}_{\alpha} = -\frac{V}{R \im \tau} \partial_\mu y(\gamma^{\mu})_{\alpha}^{\;\;\beta}\zeta'^ {a}_{\beta}\\
&\delta \bar{\lambda}'^a_{\alpha} =i(\gamma^{\mu})_{\alpha}^{\;\;\beta} {\zeta}'^{a}_{\beta}\partial_{\mu}{\phi} -\frac{iR}{2V}\partial_{\mu}(\frac{V}{R})\zeta'^ {a}_{\beta} \phi
\end{split}
\end{equation}
Comparing the above SUSY transformation with the standard form of SUSY transformation for the deformed \hk sigma model allows one to relate the fermions in the two descriptions as before:
\begin{equation}
\begin{split}
& \lambda'^a_{\alpha} =\frac{1}{\sqrt{2}} \left(\psi^{2'}_\alpha,\bar \psi_{\alpha}^ {2'} \right)\\
& \bar{\lambda}'^a_{\alpha}=\frac{1}{\sqrt{2}} \left(\psi^{1'}_\alpha, \bar \psi_{\alpha}^ {1'} \right)\\
& \zeta'^a_{\alpha}= \frac{1}{\sqrt{2}} \left(-\zeta^2_{\alpha}, -\bar{\zeta}^2_{\alpha}\right)
\end{split}
\end{equation}
Since $\zeta'^a_{\alpha} =\frac{\zeta^a_{\alpha}}{V^{1/4}}$ with $\zeta^a_{\alpha}$ being a constant spinor, the above identification immediately implies
\begin{equation}
\begin{split}
&\partial_{\mu} \zeta^E + \frac{1}{4} \frac{\partial_{\mu} V}{V}\zeta^E=0\\
&\partial_{\mu} \bar{\zeta}^E + \frac{1}{4} \frac{\partial_{\mu} V}{V}\bar{\zeta}^E=0\\
\implies & f(\varphi^0)=\frac{R}{4V}
\end{split}
\end{equation}
One can also read off $q_0$ directly from the fermionic action:
\begin{equation}
q^{A'}_{B' 0}=\frac{R}{V}\begin{pmatrix}  -1/4 & 0  \\ 0 & -3/4 \end{pmatrix},\; q^{A'}_{B' i} =0 \; (i \neq 0)
\end{equation}
Therefore, the effective $\tilde{q}^{A'}_{0B'}$ that appears in the extended \hk identity is
\begin{equation}
\begin{split}
\tilde{q}^{A'}_{B' 0}=\frac{R}{V}\begin{pmatrix}  0 & 0  \\ 0 & -1/2\end{pmatrix}
\end{split}
\end{equation}

Given the explicit forms of $\tilde{q}^{A'}_{B' 0}$ and $f(\varphi^0)$, one obtains the following nontrivial components of the connection from the extended \hk identity \eqref{eq:e-cc2}.
\begin{equation}
\begin{split}
&\Gamma^{\phi}_{0\phi}= -\frac{R}{2V}, \; \Gamma^{\bar{\phi}}_{0\bar{\phi}}= -\frac{R}{V} \\
&\Gamma^{y}_{0y}= 0, \; \Gamma^{\bar{y}}_{0\bar{y}}= \frac{R}{2V}\\
&\Gamma^{\phi}_{\phi 0}= -\frac{R}{2V}, \; \Gamma^{\bar{\phi}}_{\bar{\phi}0}= -\frac{R}{2V} \\
&\Gamma^{y}_{y0}= 0, \; \Gamma^{\bar{y}}_{\bar{y}0}= 0\\
&\Gamma^{\phi}_{00}= \frac{3}{4} \frac{R^2 \phi}{V^2}, \; \Gamma^{\bar{\phi}}_{00}= \frac{R^2 \bar{\phi}}{V^2}
\end{split} \label{conn2}
\end{equation}
Now, we can readily check whether this connection obeys the SUSY constraints \eqref{covg}-\eqref{torsionfree}, which we derived for a general deformed \hk sigma model.  For $U(1)$ SYM on NUT space, the non-trivial part of the first SUSY constraint \eqref{covg} is
\begin{equation}
\begin{split}
e^l_{EE'} \left( \partial_i g_{0l} + \partial_0 g_{il} - \partial_l g_{i0} - ({\Gamma}_{i0}^k+{\Gamma}_{0i}^k) g_{kl} +  F_{il}  \right) = 0
\end{split}
\end{equation}
From the structure of the intertwiners specified in \eqref{intertwiners} and the nonzero components of $\Gamma^k_{ij}$ in \eqref{conn2}, it is clear that there are only three nontrivial components that one needs to check, namely for $l=\bar{y}, i=y$, for $l=\bar{\phi}, i=\phi$ and for $i=0, l=\bar{\phi}$.

In the first case, we have
\begin{equation}
\begin{split}
&\partial_yg_{0\bar{y}}+\partial_0 g_{y\bar{y}}- \partial_{\bar{y}} g_{y0}-(\Gamma^k_{y0}+\Gamma^k_{0y})g_{k\bar{y}}+ F_{y\bar{y}}\\
&= \partial_0 g_{y\bar{y}}+F_{y\bar{y}}\\
&=  0,
\end{split}
\end{equation}
while the second case leads to
\begin{equation}
\begin{split}
&\partial_\phi g_{0\bar{\phi}} + \partial_0 g_{\phi\bar{\phi}} -  \partial_{\bar{\phi}} g_{\phi0} -(\Gamma^k_{\phi 0}+\Gamma^k_{0 \phi})g_{k\bar{\phi}} + F_{\phi \bar{\phi}}\\
&=\left(-\frac{1}{2} \frac{4\pi R^2 \im \tau}{V^2}\right)+\left(-\frac{4\pi R^2 \im \tau}{V^2}\right) -\left(-\frac{1}{2} \frac{4\pi R^2 \im \tau}{V^2}\right) -\left(-\frac{R}{2V}-\frac{R}{2V}\right)\frac{4\pi R\im \tau}{V}\\
&=0.
\end{split}
\end{equation}
For the third case, we get
\begin{equation}
\begin{split}
&\partial_\phi g_{0\bar{\phi}} + \partial_0 g_{0\bar{\phi}} -  \partial_{\bar{\phi}} g_{00}-2\Gamma^k_{00} g_{k\bar{\phi}}\\
&= 2(\frac{R^3 \phi \im \tau}{V^3})-\frac{R^3 \phi \im \tau}{2V^3}-2(\frac{3}{4} \frac{R^2 \phi}{V^2}\frac{R \im \tau}{V})\\
&=0.
\end{split}
\end{equation}

The connection derived in \eqref{conn2} therefore obeys the first SUSY constraint \eqref{covg}.
Finally, the second SUSY constraint \eqref{torsionfree} can be written as
\begin{align}
&e_{k2E'}(\Gamma_{i0}^k  -\Gamma_{0i}^k)=0\\
\implies & \Gamma_{i0}^y  -\Gamma_{0i}^y=0, \; \Gamma_{i0}^{\phi}  -\Gamma_{0i}^{\phi}=0\; \; \forall i
\end{align}
which is evidently satisfied in this case.

\section{SUSY variation of the \hk sigma model: 3-fermion and 5-fermion terms} \label{susy35}

The 1-fermion terms in the SUSY variation of the deformed \hk sigma model were analyzed in \S\ref{1-fermionSUSY}. In this appendix, we show that the 3-fermion terms and the 5-fermion terms also vanish such that the sigma model action is indeed invariant under the SUSY transformation \eqref{susyvar1}-- \eqref{susyvar3}. We show that vanishing of the 3-fermion terms requires that the generalization of \hk identity associated with the special form of curvature on a \hk manifold, given by \eqref{eq:hk-curvature2}, is satisfied. Similarly, vanishing of the 5-fermion terms requires that the Bianchi identity, given by \eqref{bianchi0}, is satisfied.

\subsubsection*{3-fermion terms}
 Let us consider the 3-fermion terms in the SUSY variation first.
\begin{subequations}
\begin{align}
4\pi \delta \mathcal{L} =
&-\I q_{i A'}^{E'}  \psi^{F'\sigma} \bar\zeta^F_\sigma e^i_{FF'} \bar\psi_{E'\alpha} \gamma^{\mu \alpha}_\beta (\partial_\mu \psi^{A'\beta} + q^{A'}_{B' j} \partial_\mu \varphi^j \psi^{B' \beta}) \\
&+\I\bar\psi_{\alpha A'} \gamma^{\mu \alpha}_\beta \partial_\mu ( q_{i E'}^{A'} \psi^{F'\sigma} \bar\zeta^F_\sigma e^i_{FF'} \psi^{E'\beta}) \\
&-\I\bar\psi_{\alpha A'} \gamma^{\mu \alpha}_\beta \partial_j q^{A'}_{B' i}  \psi^{E'\sigma} \bar\zeta^E_\sigma e^j_{EE'} \partial_\mu \varphi^i \psi^{B' \beta} \\
&-\I\bar\psi_{\alpha A'} \gamma^{\mu \alpha}_\beta q^{A'}_{B' i} \partial_\mu ( \psi^{E'\sigma} \bar\zeta^E_\sigma e^i_{EE'}) \psi^{B' \beta} \\
&+\I\bar\psi_{\alpha A'} \gamma^{\mu \alpha}_\beta q^{A'}_{B' i} \partial_\mu \varphi^i q_{j E'}^{B'}  \psi^{F'\sigma} \bar\zeta^F_\sigma e^j_{FF'} \psi^{E'\beta} \\
&+\I\Omega_{A'B'C'D'} \psi^{A'}_{\alpha} \gamma^{\mu\alpha}_\beta  \partial_\nu \varphi^i e_{iE}^{B'} \gamma_\sigma^{\nu \beta} \bar\zeta^{\sigma E} \psi^{C'}_\delta \gamma_{\mu \omega}^\delta \bar\psi^{D'\omega}
\end{align}
\end{subequations}
Relabeling indices this becomes $\delta \mathcal{L} = \I \bar\psi_{A'\alpha} \bar\zeta^E_\sigma \delta \mathcal{L}^{A'\alpha \sigma}_E$ where
\begin{subequations}
\begin{align}
\delta \mathcal{L}^{A'\alpha \sigma}_E =
&+ q_{i E'}^{A'}  \psi^{F'\sigma} e^i_{EF'} \gamma^{\mu \alpha}_\beta (\partial_\mu \psi^{E'\beta} + q^{E'}_{B' j} \partial_\mu \varphi^j \psi^{B' \beta}) \\
&- \gamma^{\mu \alpha}_\beta \partial_\mu (q_{i E'}^{A'} \psi^{F'\sigma} e^i_{EF'} \psi^{E'\beta}) \\
&+ \gamma^{\mu \alpha}_\beta \partial_j q^{A'}_{B' i} \psi^{E'\sigma} e^j_{EE'} \partial_\mu \varphi^i \psi^{B' \beta} \\
&+ \gamma^{\mu \alpha}_\beta q^{A'}_{B' i} \partial_\mu (\psi^{E'\sigma} e^i_{EE'}) \psi^{B' \beta} \\
&- \gamma^{\mu \alpha}_\beta q^{A'}_{B' i} \partial_\mu \varphi^i q_{j E'}^{B'} \psi^{F'\sigma} e^j_{EF'} \psi^{E'\beta} \\
&+ \Omega^{A'}_{E'B'C'} \psi^{E'}_{\tau} \gamma^{\mu\tau}_\beta \partial_\nu \varphi^i e_{iE}^{B'} \gamma^{\nu \sigma \beta} \psi^{C'}_\delta \gamma_{\mu}^{\alpha\delta}
\end{align}
\end{subequations}
which expands out to
\begin{subequations}
\begin{align}
\delta \mathcal{L}^{A'\alpha \sigma}_E =
&+ q_{i E'}^{A'}  \psi^{F'\sigma} e^i_{EF'} \gamma^{\mu \alpha}_\beta \partial_\mu \psi^{E'\beta} \\
&+ q_{i E'}^{A'}  \psi^{F'\sigma} e^i_{EF'} \gamma^{\mu \alpha}_\beta q^{E'}_{B' j} \partial_\mu \varphi^j \psi^{B' \beta} \\
&- \gamma^{\mu \alpha}_\beta \partial_\mu \varphi^j \partial_j q_{i E'}^{A'} \psi^{F'\sigma} e^i_{EF'} \psi^{E'\beta} \\
&- \gamma^{\mu \alpha}_\beta  q_{i E'}^{A'} \partial_\mu\psi^{F'\sigma} e^i_{EF'} \psi^{E'\beta} \\
&- \gamma^{\mu \alpha}_\beta q_{i E'}^{A'} \psi^{F'\sigma} \partial_\mu \varphi^j \partial_j e^i_{EF'} \psi^{E'\beta} \\
&- \gamma^{\mu \alpha}_\beta q_{i E'}^{A'} \psi^{F'\sigma} e^i_{EF'} \partial_\mu \psi^{E'\beta} \\
&+ \gamma^{\mu \alpha}_\beta \partial_j q^{A'}_{B' i} \psi^{E'\sigma} e^j_{EE'} \partial_\mu \varphi^i \psi^{B' \beta} \\
&+ \gamma^{\mu \alpha}_\beta q^{A'}_{B' i} \partial_\mu \psi^{E'\sigma} e^i_{EE'} \psi^{B' \beta} \\
&+ \gamma^{\mu \alpha}_\beta q^{A'}_{B' i} \psi^{E'\sigma} \partial_\mu \varphi^j \partial_j e^i_{EE'} \psi^{B' \beta} \\
&- \gamma^{\mu \alpha}_\beta q^{A'}_{B' i} \partial_\mu \varphi^i q_{j E'}^{B'} \psi^{F'\sigma} e^j_{EF'} \psi^{E'\beta} \\
&+ \Omega^{A'}_{E'B'C'} \psi^{E'}_{\tau} \gamma^{\mu\tau}_\beta \partial_\nu \varphi^i e_{iE}^{B'} \gamma^{\nu \sigma \beta} \psi^{C'}_\delta \gamma_{\mu}^{\alpha\delta}
\end{align}
\end{subequations}

Now we may divide this into the terms involving derivatives of fermions and those involving
derivatives of bosons.  First, the terms with derivatives of fermions add up to zero:
\begin{subequations}
\begin{align}
\delta \mathcal{L}^{A'\alpha \sigma}_E =
&+ q_{i E'}^{A'}  \psi^{F'\sigma} e^i_{EF'} \gamma^{\mu \alpha}_\beta \partial_\mu \psi^{E'\beta} \\
&- \gamma^{\mu \alpha}_\beta  q_{i E'}^{A'} \partial_\mu\psi^{F'\sigma} e^i_{EF'} \psi^{E'\beta} \\
&- \gamma^{\mu \alpha}_\beta q_{i E'}^{A'} \psi^{F'\sigma} e^i_{EF'} \partial_\mu \psi^{E'\beta} \\
&+ \gamma^{\mu \alpha}_\beta q^{A'}_{B' i} \partial_\mu \psi^{E'\sigma} e^i_{EE'} \psi^{B' \beta}\\
&=0.
\end{align}
\end{subequations}

Next, the derivatives of bosons:
\begin{subequations}
\begin{align}
\delta \mathcal{L}^{A'\alpha \sigma}_E =
&+ q_{i E'}^{A'}  \psi^{F'\sigma} e^i_{EF'} \gamma^{\mu \alpha}_\beta q^{E'}_{B' j} \partial_\mu \varphi^j \psi^{B' \beta} \\
&- \gamma^{\mu \alpha}_\beta \partial_\mu \varphi^j \partial_j q_{i E'}^{A'} \psi^{F'\sigma} e^i_{EF'} \psi^{E'\beta} \\
&- \gamma^{\mu \alpha}_\beta q_{i E'}^{A'} \psi^{F'\sigma} \partial_\mu \varphi^j \partial_j e^i_{EF'} \psi^{E'\beta} \\
&+ \gamma^{\mu \alpha}_\beta \partial_j q^{A'}_{B' i} \psi^{E'\sigma} e^j_{EE'} \partial_\mu \varphi^i \psi^{B' \beta} \\
&+ \gamma^{\mu \alpha}_\beta q^{A'}_{B' i} \psi^{E'\sigma} \partial_\mu \varphi^j \partial_j e^i_{EE'} \psi^{B' \beta} \\
&- \gamma^{\mu \alpha}_\beta q^{A'}_{B' i} \partial_\mu \varphi^i q_{j E'}^{B'} \psi^{F'\sigma} e^j_{EF'} \psi^{E'\beta} \\
&+ \Omega^{A'}_{E'B'C'} \psi^{E'}_{\tau} \gamma^{\mu\tau}_\beta \partial_\nu \varphi^i e_{iE}^{B'} \gamma^{\nu \sigma \beta} \psi^{C'}_\delta \gamma_{\mu}^{\alpha\delta}
\end{align}
\end{subequations}
which rearranges to $\delta \mathcal{L}^{A'\alpha\sigma}_E = \delta \mathcal{L}^{A'\alpha\sigma\mu}_{Ei} \partial_\mu \varphi^i$ with
\begin{subequations}
\begin{align}
\delta \mathcal{L}^{A'\alpha \sigma}_{Ei} =
&+ \gamma^{\mu \alpha}_\beta q_{j B'}^{A'} q^{B'}_{E' i}  e^j_{EF'} \psi^{F'\sigma} \psi^{E'\beta} \\
&- \gamma^{\mu \alpha}_\beta \partial_i q_{j E'}^{A'}  e^j_{EF'}    \psi^{F'\sigma} \psi^{E'\beta} \\
&- \gamma^{\mu \alpha}_\beta q_{j E'}^{A'}  \partial_i e^j_{EF'}    \psi^{F'\sigma} \psi^{E'\beta} \\
&+ \gamma^{\mu \alpha}_\beta \partial_j q^{A'}_{E' i}  e^j_{EF'}    \psi^{F'\sigma} \psi^{E'\beta} \\
&+ \gamma^{\mu \alpha}_\beta q^{A'}_{E' j} \partial_i e^j_{EF'}     \psi^{F'\sigma} \psi^{E'\beta} \\
&- \gamma^{\mu \alpha}_\beta q^{A'}_{B' i} q_{j E'}^{B'} e^j_{EF'}   \psi^{F'\sigma} \psi^{E'\beta} \\
&+ \Omega^{A'}_{E'B'C'} e_{iE}^{B'} \gamma^{\nu\tau}_\beta \gamma_{\nu}^{\alpha\delta}\gamma^{\mu \sigma \beta} \psi^{E'}_{\tau} \psi^{C'}_\delta
\end{align}
\end{subequations}
i.e.
\begin{subequations}
\begin{align}
\delta \mathcal{L}^{A'\alpha \sigma\mu}_{Ei} =
&+ \gamma^{\mu \alpha}_\beta (+q_{j B'}^{A'} q^{B'}_{E' i} - q^{A'}_{B' i} q_{j E'}^{B'} - \partial_i q_{j E'}^{A'} + \partial_j q^{A'}_{E' i})  e^j_{EF'} \psi^{F'\sigma} \psi^{E'\beta} \\
&+ \Omega^{A'}_{E'B'C'} e_{iE}^{B'} \gamma^{\nu\tau}_\beta \gamma_{\nu}^{\alpha\delta}\gamma^{\mu \sigma \beta} \psi^{E'}_{\tau} \psi^{C'}_\delta
\end{align}
\end{subequations}
which we may relabel to $\delta \mathcal{L}^{A'\alpha \sigma\mu}_{Ei} = \delta \mathcal{L}^{A'\alpha \sigma\mu}_{EiE'F'\beta\tau} \psi^{F'\tau} \psi^{E'\beta}$, where
\begin{subequations}
\begin{align}
\delta \mathcal{L}^{A'\alpha \sigma\mu}_{EiE'F'\beta\tau} =
&+ \gamma^{\mu \alpha}_\beta (+q_{j B'}^{A'} q^{B'}_{E' i} - q^{A'}_{B' i} q_{j E'}^{B'} - \partial_i q_{j E'}^{A'} + \partial_j q^{A'}_{E' i})  e^j_{EF'} \delta^\sigma_\tau \\
&+ \Omega^{A'}_{F'B'E'} \gamma_{\nu\beta}^{\alpha}  (\gamma^{\nu} \gamma^\mu)^\sigma_{\tau} e_{iE}^{B'}
\end{align}
\end{subequations}

Finally using the gamma matrix identity: $\gamma_{\nu\beta}^{\alpha} ((\gamma^{\nu} \gamma^\mu)^\sigma_\tau + g^{\mu \nu} \delta^\sigma_{\tau}) - (\beta \leftrightarrow \tau) = 0$,
one can modify the variation above to
\begin{equation}
\delta \mathcal{L}^{A'\alpha \sigma\mu}_{EiE'F'\beta\tau} =
\gamma^{\mu \alpha}_\beta \delta^\sigma_\tau \left( (q_{j B'}^{A'} q^{B'}_{E' i} - q^{A'}_{B' i} q_{j E'}^{B'} - \partial_i q_{j E'}^{A'} + \partial_j q^{A'}_{E' i}) e^j_{EF'} - \Omega^{A'}_{F'B'E'} e_{iE}^{B'} \right)
\end{equation}
which indeed vanishes according to the extended \hk identity \eqref{eq:hk-curvature2}.

\subsection*{5-fermion terms}

Now, consider the 5-fermion terms in the SUSY variation.

\begin{subequations}
\begin{align}
4\pi \delta \mathcal{L} =
&- \half \partial_i \Omega_{A'B'C'D'}  \psi^{E'\sigma} \bar\zeta^E_\sigma e^i_{EE'} \psi^{A'}_{\alpha} \gamma^{\mu\alpha}_\beta \bar\psi^{B'\beta} \psi^{C'}_\delta \gamma_{\mu \omega}^\delta \bar\psi^{D'\omega} \\
&+ \Omega_{A'B'C'D'}  q_{i E'}^{A'} \psi^{F'\sigma} \bar\zeta^F_\sigma e^i_{FF'}  \psi^{E'}_{\alpha} \gamma^{\mu\alpha}_\beta \bar\psi^{B'\beta} \psi^{C'}_\delta \gamma_{\mu \omega}^\delta \bar\psi^{D'\omega} \\
&- \Omega_{A'B'C'D'} \psi^{A'}_{\alpha} \gamma^{\mu\alpha}_\beta  q_{i}^{E'B'}  \psi^{F'\sigma} \bar\zeta^F_\sigma e^i_{FF'} \bar\psi_{E'}^{\beta} \psi^{C'}_\delta \gamma_{\mu \omega}^\delta \bar\psi^{D'\omega}
\end{align}
\end{subequations}
which we may reorder/relabel into
\begin{subequations}
\begin{align}
4\pi \delta \mathcal{L} =
&- \half \partial_i \Omega_{A'B'C'D'} e^i_{FF'} \gamma^{\mu\alpha}_\beta \gamma_{\mu \omega}^\delta  \bar\zeta^F_\sigma \psi^{F'\sigma}   \psi^{A'}_{\alpha} \psi^{C'}_\delta \bar\psi^{B'\beta}  \bar\psi^{D'\omega} \\
&+ \Omega_{E'B'C'D'} q_{i A'}^{E'} e^i_{FF'} \gamma^{\mu\alpha}_\beta \gamma_{\mu \omega}^\delta  \bar\zeta^F_\sigma \psi^{F'\sigma}  \psi^{A'}_\alpha \psi^{C'}_\delta \bar\psi^{B'\beta} \bar\psi^{D'\omega} \\
&+ \Omega_{A'E'C'D'} q_{i B'}^{E'} e^i_{FF'} \gamma^{\mu\alpha}_\beta \gamma_{\mu \omega}^\delta \bar\zeta^F_\sigma \psi^{F'\sigma} \psi^{A'}_{\alpha}  \psi^{C'}_\delta \bar\psi^{B'\beta}  \bar\psi^{D'\omega}
\end{align}
\end{subequations}
or equivalently, using the symmetry under exchange $\alpha \beta \leftrightarrow \delta \omega$,
\begin{subequations}
\begin{align}
4 \pi \delta \mathcal{L}=
&- \half \partial_i \Omega_{A'B'C'D'} e^i_{FF'} \gamma^{\mu\alpha}_\beta \gamma_{\mu \omega}^\delta  \bar\zeta^F_\sigma \psi^{F'\sigma}   \psi^{A'}_{\alpha} \psi^{C'}_\delta \bar\psi^{B'\beta}  \bar\psi^{D'\omega} \\
&+ \half \Omega_{E'B'C'D'} q_{i A'}^{E'} e^i_{FF'} \gamma^{\mu\alpha}_\beta \gamma_{\mu \omega}^\delta  \bar\zeta^F_\sigma \psi^{F'\sigma}  \psi^{A'}_\alpha \psi^{C'}_\delta \bar\psi^{B'\beta} \bar\psi^{D'\omega} \\
&+ \half \Omega_{E'D'A'B'} q_{i C'}^{E'} e^i_{FF'} \gamma^{\mu\alpha}_\beta \gamma_{\mu \omega}^\delta  \bar\zeta^F_\sigma \psi^{F'\sigma}  \psi^{A'}_\alpha \psi^{C'}_\delta \bar\psi^{B'\beta} \bar\psi^{D'\omega} \\
&+ \half \Omega_{A'E'C'D'} q_{i B'}^{E'} e^i_{FF'} \gamma^{\mu\alpha}_\beta \gamma_{\mu \omega}^\delta \bar\zeta^F_\sigma \psi^{F'\sigma} \psi^{A'}_{\alpha}  \psi^{C'}_\delta \bar\psi^{B'\beta}  \bar\psi^{D'\omega} \\
&+ \half \Omega_{C'E'A'B'} q_{i D'}^{E'} e^i_{FF'} \gamma^{\mu\alpha}_\beta \gamma_{\mu \omega}^\delta \bar\zeta^F_\sigma \psi^{F'\sigma} \psi^{A'}_{\alpha}  \psi^{C'}_\delta \bar\psi^{B'\beta}  \bar\psi^{D'\omega}\\
=&-\half B_{iA'B'C'D'} e^i_{FF'} \gamma^{\mu\alpha}_\beta \gamma_{\mu \omega}^\delta  \bar\zeta^F_\sigma \psi^{F'\sigma}   \psi^{A'}_{\alpha} \psi^{C'}_\delta \bar\psi^{B'\beta}  \bar\psi^{D'\omega}
\end{align}
\end{subequations}
where in the last step we have used the definition of $B_{i A'B'C'D'}$ in \eqref{bianchi0}
\begin{equation}
B_{i A'B'C'D'} = \partial_i \Omega_{A'B'C'D'} - q_{i A'}^{E'} \Omega_{E'B'C'D'} - q_{i B'}^{E'} \Omega_{A'E'C'D'} - q_{i C'}^{E'} \Omega_{A'B'E'D'} - q_{i D'}^{E'} \Omega_{A'B'C'E'} \nonumber
\end{equation}

Note that $B_{i A'B'C'D'}$ is completely symmetric in the $Sp(r)$ indices - this follows from the symmetry property of $\Omega_{A'B'C'D'}$  and the above definition.

Now, using the identity $\gamma^{\mu\alpha}_\beta \gamma_{\mu \omega}^\delta= 2\delta^{\alpha}_{\omega}\delta^{\delta}_{\beta}- \delta^{\alpha}_{\beta}\delta^{\delta}_{\omega}$, one can show that
\begin{equation}
\begin{split}
&\gamma^{\mu\alpha}_\beta \gamma_{\mu \omega}^\delta  \bar\zeta^F_\sigma \psi^{F'\sigma}   \psi^{A'}_{\alpha} \psi^{C'}_\delta \bar\psi^{B'\beta}  \bar\psi^{D'\omega}= - \bar\zeta^F_\sigma \psi^{A'\sigma}  \psi^{F'}_{\alpha} \bar\psi^{D'\alpha} \psi^{C'}_\delta \bar\psi^{B'\delta}+\bar\zeta^F_\sigma \psi^{F'\sigma}  \psi^{A'}_{\alpha} \bar\psi^{B'\alpha} \psi^{C'}_\delta \bar\psi^{D'\delta}
\end{split}
\end{equation}
Therefore, the 5-fermion term reduces to
\begin{equation}
\begin{split}
\delta \mathcal{L}=& -\half B_{iA'B'C'D'} e^i_{FF'} \left(- \bar\zeta^F_\sigma \psi^{A'\sigma}  \psi^{F'}_{\alpha} \bar\psi^{D'\alpha} \psi^{C'}_\delta \bar\psi^{B'\delta}+\bar\zeta^F_\sigma \psi^{F'\sigma}  \psi^{A'}_{\alpha} \bar\psi^{B'\alpha} \psi^{C'}_\delta \bar\psi^{D'\delta}\right)\\
=&\left(B_{iF'B'C'D'} e^i_{FA'}-B_{iA'B'C'D'} e^i_{FF'}\right)\bar\zeta^F_\sigma \psi^{F'\sigma}  \psi^{A'}_{\alpha} \bar\psi^{B'\alpha} \psi^{C'}_\delta \bar\psi^{D'\delta}
\end{split}
\end{equation}

Now recall the Bianchi identity from \eqref{bianchi1},
\begin{equation}
B_{i A' B' C' D'} e^i_{FF'} - B_{i F'B'C'D'} e^i_{F A'} = 0. \nonumber
\end{equation}

Using the Bianchi identity, the 5-fermion term in the SUSY variation evidently vanishes.

\section{6D spinors and \tops{$\mathcal{N}=1$}{N=1} SYM in 6D} \label{6Dspinor}

In this section we explain our conventions regarding 6D spinors and provide a few more details about the SUSY transformation of fields in 6D $\mathcal{N}=1$ SYM.

The Lagrangian of the 6D theory is
\begin{equation}
 \mathcal{L}_6 = \frac{1}{g^2_6} \left[\frac{1}{2} F_{MN} F^{MN} + \bar{\psi}_a \Gamma^M \partial_M \psi^a+\frac{1}{2} D_{ab}D^{ab}\right] \label{6DSYMflat}
\end{equation}
where we choose the following metric on the flat space $\eta_{MN}=\left(-1,1,1,1,1,1\right)$. The fermionic field $\psi_a$ is a symplectic Majorana-Weyl spinor which transforms as a doublet of the $SU(2)$ R-symmetry. A $SO(1,5)$ spinor obeys the Weyl condition and is conjugate to self but does not obey the standard "Majorana" condition. However, when combined with the $SU(2)_R$ symmetry, one can have a modified reality condition on these spinors - the "symplectic Majorana" condition.
\begin{equation}
\begin{split}
&\Gamma^7 \psi_a =\psi_a \; (\mbox{Weyl condition})\\
&\psi^{a,\; T}C^{-}_6 \epsilon_{ab}= (\psi^{b\;})^{\dagger} \I\Gamma^0 \equiv \bar{\psi}^b  \;(\mbox{Symplectic Majorana condition})
\end{split}
\end{equation}
where $\Gamma^7$ is the chirality matrix in 6D defined as $\Gamma^7=-\Gamma^0 \Gamma^1 \Gamma^2 \Gamma^3 \Gamma^4 \Gamma^5$. $C^{-}_6$ is the 6D charge conjugation obeying $C^{-}_6 \Gamma^{\mu} (C^{-}_6)^{-1}=-(\Gamma^{\mu})^T$. \\

\subsubsection*{SUSY transformation}
The action in equation (\ref{6DSYMflat}) is invariant under the following SUSY transformation rules:
\begin{equation}
\begin{split}
&\delta A_M=\frac{1}{2}(\bar{\zeta}_a \Gamma_M \psi^a-\bar{\psi}_a \Gamma_M \zeta^a)= -\bar{\psi}_a \Gamma_M \zeta^a \;(\mbox{since} \; \;\bar{\zeta}_a \Gamma_M \psi^a=-\bar{\psi}_a \Gamma_M \zeta^a)\\
 &\delta \psi_a=-\frac{1}{2} F_{MN} \Gamma^{MN} \zeta_a - D_{ab} \zeta^b, \;  \delta \bar{\psi}_a= \frac{1}{2} \bar{\zeta}_a \Gamma^{MN} F_{MN} - D_{ab} \bar{\zeta}^b\\
 &\delta D_{ab}= \bar{\zeta}_a \Gamma^M \partial_M \psi_b + \bar{\zeta}_b \Gamma^M \partial_M \psi_a
\end{split}\label{6Dsusyflat_App}
\end{equation}
Note that the SUSY parameter $\zeta_a$ is a Grassman-odd symplectic Majorana-Weyl spinor and  a solution of the Killing spinor equation on $\mathbb{R}^3 \times S^1\times T^{1,1}$ :
\begin{equation}
\partial_M \zeta^a=0 \implies \zeta^a =\mbox{constant}
\end{equation}
 In equation (\ref{6Dsusyflat_App}), we used that $\bar{\psi}_a \Gamma_M \zeta^a=-\bar{\zeta}_a \Gamma_M \psi^a$ which follows from the general relation involving 6D spinors
\begin{equation}
\bar{\psi}_a \Gamma_{N_1\;N_2 \ldots N_n} \zeta^a=(-1)^n \bar{\zeta}_a \Gamma_{N_n\;N_{n-1} \ldots N_1} \psi^a
\end{equation}
The SUSY variation for $\bar{\psi}_a$ can be obtained as follows:
\begin{equation}
\begin{split}
\delta \bar{\psi}_a= & \delta {\psi}^{\dagger}_a i \Gamma_0 =-\frac{\I}{2} F_{MN} (\Gamma^{MN} \zeta_a)^{\dagger} \Gamma_0
=-\frac{\I}{2} F_{MN} \zeta_a^{\dagger} (\Gamma^{MN} )^{\dagger} \Gamma_0 \\
=&-\frac{1}{2} F_{MN} \bar{\zeta}_a \Gamma^{NM} \;(\mbox{using}\; \Gamma^{MN \dagger} \Gamma^0=\Gamma^0 \Gamma^{MN})= \frac{1}{2} F_{MN} \bar{\zeta}_a \Gamma^{MN}
\end{split} \nonumber
\end{equation}

\subsubsection*{SUSY invariance of the action}

The variation of the bosonic part of the Lagrangian is
\begin{equation}
\begin{split}
\delta \mathcal{L}^b_6= &\frac{1}{g^2_6} \delta (\frac{1}{2} F_{MN} F^{MN})+\frac{1}{g^2_6} \delta (\frac{1}{2} D_{ab} D^{ab})\\
=&-\frac{1}{g^2_6} F^{MN} \left(\partial_M \bar{\psi}_a \Gamma_N \zeta^a -\partial_N \bar{\psi}_a \Gamma_M \zeta^a\right)+\frac{1}{g^2_6} D^{ab} \left(\bar{\zeta}_a \Gamma^M \partial_M \psi_b + \bar{\zeta}_b \Gamma^M \partial_M \psi_a \right)\\
=&-\frac{1}{g^2_6} F^{MN} \left(\partial_M \bar{\psi}_a \Gamma_N \zeta^a -\partial_N \bar{\psi}_a \Gamma_M \zeta^a\right)+\frac{2}{g^2_6} D^{ab} \bar{\zeta}_a \Gamma^M \partial_M \psi_b\; \;(D_{ab}\; \mbox{is symmetric})
\end{split} \label{bosvar6D}
\end{equation}
The variation of the fermionic terms in the Lagrangian is
\begin{equation}
\begin{split}
\delta \mathcal{L}^f_6= &\frac{1}{g^2_6} \delta (\bar{\psi}_a \Gamma^M \partial_M \psi^a)\\
=&\frac{1}{g^2_6} \Big(\frac{1}{2} F_{MN} \bar{\zeta}_a \Gamma^{MN}\Gamma^P \partial_P \psi^a-D_{ab}\bar{\zeta}^b\Gamma^M \partial_M \psi^a -\frac{1}{2} \bar{\psi}_a\Gamma^P \partial_P(F_{MN}\Gamma^{MN} \zeta^a)\\
&-\bar{\psi}_a \Gamma^M \partial_M(D^a_{\;\;b} \zeta^b) \Big)\\
\xrightarrow[\mbox{by parts}]{\mbox{Integration}}&\frac{1}{g^2_6}\Big(\frac{1}{2} F_{MN} \bar{\zeta}_a \Gamma^{MN}\Gamma^P \partial_P \psi^a+\frac{1}{2} F_{MN} \partial_P \bar{\psi}_a\Gamma^P \Gamma^{MN} \zeta^a-D_{ab}\bar{\zeta}^b\Gamma^M \partial_M \psi^a\\
&- D_{ab} \partial_M \bar{\psi}^a \Gamma^M  \zeta^b\Big)\\
=&\frac{1}{g^2_6} F^{MN} \left(\partial_M\bar{\psi}_a \Gamma_{N}  \zeta^a-\partial_N\bar{\psi}_a \Gamma_{M}  \zeta^a-2D_{ab}\bar{\zeta}^b\Gamma^M \partial_M \psi^a\right)\\
\end{split} \label{fermvar6D}
\end{equation}
To obtain the final equation one needs to use the Bianchi identity for the gauge field i.e. $\de F=0$, in addition to the following identities involving gamma matrices:
\begin{equation}
\begin{split}
\Gamma^{MN}\Gamma^{P}=&\Gamma^{PMN} -\eta^{MP}\Gamma^{N}+\eta^{NP} \Gamma^{M} \\
\Gamma^{P}\Gamma^{MN}=& \Gamma^{PMN} +\eta^{PM}\Gamma^{N} -\eta^{PN}\Gamma^{M}
\end{split}
\end{equation}

Therefore, from \eqref{bosvar6D} and \eqref{fermvar6D}, we have
\begin{equation}
\delta_{SUSY} S_6 = 0.
\end{equation}

\subsubsection*{Closure of the SUSY algebra}

Since we took the SUSY parameter $\zeta^a$ to be Grassmann-odd, the operator $\delta_{SUSY}$ acts on the fields as a bosonic operator. Therefore, one needs to compute the action of the commutator of two such operators on the fields to check the closure of the SUSY algebra.
\begin{equation}
\begin{split}
&\Big[\delta_{\zeta'},\delta_{\zeta}\Big]A_M=2\bar{\zeta'}_a \Gamma^N \zeta^a F_{MN} \\
&\Big[\delta_{\zeta'},\delta_{\zeta}\Big]\psi^a= 2 \bar{\zeta'}_a \Gamma^N \zeta^a \partial_N \psi^a\\
&\Big[\delta_{\zeta'},\delta_{\zeta}\Big]D_{ab}= 2 \bar{\zeta'}_a \Gamma^N \zeta^a \partial_N D_{ab}
\end{split}
\end{equation}
The action of two successive SUSY transformation produces a translation with the parameter $v^N=2\bar{\zeta'}_a \Gamma^N \zeta^a$.

\section{4D and 3D spinors}\label{4D/3Dspinors}

In this section, we spell out the connection between spinors in 6D Minkowski space and spinors in 4D and 3D Euclidean space.
To go back and forth between the 6D and the 4D description, we choose the following representation of the 6D gamma matrices:
\begin{equation}
\begin{split}
&\Gamma^M = \{-\I \sigma_2 \otimes \gamma^5 \;,\textbf{1}_{2\times 2} \otimes \gamma^{m} ,\;\sigma_1 \otimes \gamma^5\} \; (m=1,2,3,4)\\
&\Gamma^7 = \sigma_3 \otimes \gamma^5
\end{split} \label{gamma6dto4d}
\end{equation}
where $\gamma^m$ are the 4D gamma matrices and $\gamma^5$ is the 4D chirality operator. In this representation, we may write any 6D Dirac spinor $\psi^a$ as the doublet $\psi^a=\left(\begin{matrix}
\psi^a_1\\
\psi^a_2\\
\end{matrix}\right)$, where each of the entries is a four-component complex spinor.

The Weyl condition $\Gamma^7 \psi_a =\psi_a$ implies
\begin{equation}
\begin{split}
&\gamma^5  \psi^a_1=\psi^a_1 \implies \psi^a_1=\left(\begin{matrix}
\lambda^{\alpha, \;a}\\
0\\
\end{matrix}\right)\\
&\gamma^5  \psi^a_2=-\psi^a_2 \implies \psi^a_2=\left(\begin{matrix}
0\\
\bar{\lambda}_{\dot{\alpha}} ^{a}\\
\end{matrix}\right)\\
\end{split}\label{spinor6dto4d}
\end{equation}
The symplectic Majorana condition, on the other hand, implies the following reality condition on $\lambda^{\alpha, \;a}$ and $\bar{\lambda}_{\dot{\alpha}}^{a}$:
\begin{equation}
\begin{split}
&(\lambda^{\alpha, a})^{\dagger} = \I \lambda^{\beta, b} \epsilon_{\beta \alpha} \epsilon_{ba}\\
&(\bar{\lambda}_{\dot{\alpha}}^{a})^{\dagger} =-\I\bar{\lambda}_{\dot{\beta}}^{b} \epsilon_{ba} \epsilon^{\dot{\beta}\dot{\alpha}}\\
\end{split}\label{reality4d}
\end{equation}
From equations (\ref{gamma6dto4d})-(\ref{reality4d}), one can now write the fermionic terms in the
Lagrangian in terms of the 4D spinors:
\begin{equation}
\begin{split}
\int_{\mathbb{R}^3 \times S^1} \de^4x\; \bar{\psi}_a \Gamma^m \partial_m \psi^a=&\int_{\mathbb{R}^3 \times S^1} \de^4x\; \Big[\bar{\lambda}^{\dagger}_a \bar{\sigma}^m\partial_m\lambda^a+{\lambda}^{\dagger}_a {\sigma}^m\partial_m \bar{\lambda}^a\Big]\\
=&\int_{\mathbb{R}^3 \times S^1} \de^4x \left(-2\I \bar{\lambda}_a \bar{\sigma}^m\partial_m\lambda^a\right)
\end{split}
\end{equation}
Similarly, the SUSY parameter $\zeta^a=\left(\begin{matrix}
\zeta^a_1\\
\zeta^a_2\\
\end{matrix}\right)$ can be decomposed as $\zeta^a_1=\left(\begin{matrix}
\zeta^{a,\alpha}\\
0\\
\end{matrix}\right)$, $\zeta^a_2=\left(\begin{matrix}
0\\
\bar{\zeta}^{a}_{\dot{\alpha}}\\
\end{matrix}\right)$ so that the SUSY transformation can also be written in terms of the 4D spinors. For example, $\delta A_m=-\bar{\psi}_a \Gamma_m \zeta^a=\I\left(\bar{\zeta}_a \bar{\sigma}^m \lambda^a - {\zeta}_a {\sigma}^m \bar{\lambda}^a\right)$. The remaining rules can be derived similarly.

Now consider further dimensional reduction of 4D spinors to 3D spinors.

 If $(x_1,x_2,x_3,x_4)$ denote local coordinates on the manifold $M_4$, then we treat $x_2$ as the circle direction. The basic rules for writing the Lagrangian and the SUSY transformation for the fields can be summarized as follows:
\begin{equation}
\begin{split}
&({\sigma}^m)^{\alpha\dot{\beta}} \to (\gamma^i)^{\alpha\beta}=(\gamma^i)^{\beta\alpha}=(\sigma_1,\sigma_3,i)\; (m \neq 2) \\
&(\bar{\sigma}^m)_{\dot{\alpha}{\beta}} \to  (\gamma^i)_{\alpha\beta}=(\gamma^i)_{\beta\alpha}=(\sigma_1,\sigma_3,i)\;(m \neq 2)\\
&({\sigma}^2)^{\alpha\dot{\beta}} \to  \I \epsilon^{\alpha\beta},\; (\bar{\sigma}^2)_{\dot{\alpha}{\beta}} \to i\epsilon_{\alpha\beta}\\
&\lambda^{a,\alpha} \to  \lambda^{a,\alpha}, \;\bar{\lambda}^{a}_{\dot{\alpha}} \to \bar{\lambda}^{a}_{{\alpha}}
\end{split}
\end{equation}
where we normalize the antisymmetric tensor $\epsilon_{\alpha\beta}$ as $\epsilon_{12}=\epsilon^{21}=1$.
The matrices $\gamma_i :=(\gamma_i)_{\alpha}^{\;\;\beta}=(-\sigma_3,\sigma_1,-\sigma_2)$ obey the following identities
\begin{equation}
\begin{split}
&\{\gamma_i,\gamma_j\}=\delta_{ij} \textbf{1} \;(\mbox{Clifford algebra})\\
&\gamma_i \gamma_j=\delta_{ij} \textbf{1} + i \epsilon_{ijk}\gamma_k \; (\mbox{Product Identity})
\end{split}
\end{equation}
The fermionic Lagrangian on dimensional reduction can therefore be rewritten as
\begin{equation}
\int_{\mathbb{R}^3 \times S^1} \de^4x \left(-2\I \bar{\lambda}_a \bar{\sigma}^m\partial_m\lambda^a\right)=-2\I\int_{\mathbb{R}^3} \de^3x\; \bar{\lambda}^{\alpha}_a ({\gamma}^i)_{\alpha}^{\;\;\beta}\partial_i\lambda^a_{\beta}
\end{equation}
Similarly, the SUSY transformation rules can be rewritten in terms of the 3D spinors,
using the above rules.

\section{Killing spinor on \tops{$TN_4 \times T^{1,1}$}{TN4 x T1,1}}\label{KSonNUT}

The metric on the space $TN_4 \times T^{1,1}$ is
\begin{equation}
\de s^2= V(r) \de r^2 + V(r) r^2 (\de\theta^2 + \sin^2{\theta} \de\phi^2) + \frac{R^2}{V}(\de\chi -\cos{\theta}d\phi)^2 + \de u^2 - \de v^2 \label{NUTmetric}
\end{equation}
with $\theta \in [0,\pi], \phi \in [0,2\pi], \chi \in [0,4\pi]$ and $V(r)=1+R/r $.
The isometry group of $TN_4$ (a \hk manifold of quaternionic dimension 1) is $U(1) \times SU(2)$ and the corresponding Killing vectors are:
\begin{align}
&X_0 = \frac{\partial}{\partial \chi}\\
&X_1=-\sin{\phi}\frac{\partial}{\partial \theta} -\cos{\phi}\cot{\theta}\frac{\partial}{\partial \phi} -\frac{\cos{\phi}}{\sin{\theta}}\frac{\partial}{\partial \chi}\\
&X_2=-\cos{\phi}\frac{\partial}{\partial \theta} +\sin{\phi}\cot{\theta}\frac{\partial}{\partial \phi} +\frac{\sin{\phi}}{\sin{\theta}}\frac{\partial}{\partial \chi}\\
&X_3=\frac{\partial}{\partial \phi}
\end{align}
The $X_i$s satisfy the $su(2)$ Lie algebra while $X_0$ is the Killing vector which generates the $U(1)$ isometry.

We will solve the Killing spinor equation on NUT space in a gauge (i.e. for a certain choice of veirbeins) where the invariance of the Killing spinor under the $U(1)$ isometry becomes manifest. Therefore, let
\begin{align}
&e^1=\sqrt{V} \de r,\\
&e^2=r\sqrt{V} \de \theta,\\
&e^3=r\sqrt{V}\sin{\theta} \de \phi, \\
&e^4=\frac{R}{V^{1/2}}(\de \chi - \cos{\theta} \de \phi), \\
&e^5=\de u, \\
&e^6=\de v
\end{align}
The independent, nonzero spin connections are then
\begin{align}
&\omega^{21}=\frac{1}{rV} \frac{\de(r\sqrt{V})}{\de r} e^2=(1-\frac{R}{2rV})\de\theta\\
&\omega^{31}=\frac{1}{rV} \frac{\de(r\sqrt{V})}{\de r} e^3=(1-\frac{R}{2rV})\sin{\theta} \de\phi\\
&\omega^{41}=\frac{-1}{2V^{3/2}} \frac{\de V}{\de r} e^4=\frac{R^2}{2r^2V^2}(\de\chi - \cos{\theta} \de\phi)\\
&\omega^{24}=\frac{-R}{2r^2V^{3/2}} e^3=\frac{-R}{2rV}\sin{\theta} \de\phi\\
&\omega^{34}=\frac{R}{2r^2V^{3/2}} e^1=\frac{R}{2rV}\de\theta\\
&\omega^{23}=-\frac{R}{2r^2V^{3/2}} e^4 - \frac{\cos{\theta}}{r\sqrt{V}\sin{\theta}}e^3=-\frac{R^2}{2r^2V^2}\de\chi -(1-\frac{R^2}{2r^2V^2})\cos{\theta} \de\phi
\end{align}

\subsection*{Solution for the Killing Spinor}
The Killing spinor equation on the manifold $TN_4 \times T^{1,1}$ is
\begin{equation}
D_{M}\zeta=\partial_{M} \zeta + \frac{1}{4}\omega^{ab}_{M}\Gamma_{ab}\zeta=0
\end{equation}
In terms of the local coordinates of \eqref{NUTmetric}, the components of the Killing spinor equation as follows
\begin{equation}
\begin{split}
&\partial_r \zeta^a = 0, \\
&\partial_{\theta} \zeta^a - \frac{1}{2} \Gamma_{12} \zeta^a + \frac{R}{4rV} \left(\Gamma_{12}+ \Gamma_{34}\right)\zeta^a=0,\\
&\partial_{\phi} \zeta^a + \frac{1}{2} \sin{\theta}\Gamma_{31} \zeta^a -\frac{1}{2} \cos{\theta}\Gamma_{23} \zeta^a - \frac{R\sin{\theta}}{4rV} \left(\Gamma_{31}+ \Gamma_{24}\right)\zeta^a -\frac{R^2\cos{\theta}}{4r^2V^2} \left(\Gamma_{41}-\Gamma_{23}\right)\zeta^a=0,\\
&\partial_{\chi} \zeta^a + \frac{R^2}{4r^2V^2} \left(\Gamma_{41}-\Gamma_{23}\right)\zeta^a=0,\\
& \partial_{u} \zeta^a =0,\\
&\partial_{v} \zeta^a =0.
\end{split}
\end{equation}
The terms dependent on the radial coordinate drop off if we choose the spinor such that $\left(\textbf{1} - \Gamma_1\Gamma_2\Gamma_3\Gamma_4\right)\zeta^a=0$. Therefore, for any given representation of the gamma matrices $\Gamma_{ab}$, the solution of this equation is,
\begin{equation}
\boxed{\zeta= e^{\frac{\theta}{2} \Gamma_{12}}e^{\frac{\phi}{2} \Gamma_{23}} \zeta_0,\;\;(\textbf{1}-\Gamma_1\Gamma_2 \Gamma_3 \Gamma_4) \zeta_0= 0} \label{TN-KS}
\end{equation}
 With this projection condition imposed, the Killing spinor equation is clearly the same as that for $\mathbb{R}^3  \times S^1\times T^{1,1}$ written in spherical polar coordinates on $\mathbb{R}^3$ for which we know the Killing spinor to be simply a constant spinor. Therefore, for a particular choice of vierbeins (compatible with Cartesian coordinates on $\mathbb{R}^3$), the Killing spinor on $TN_4 \times T^{1,1}$ is simply a constant spinor obeying the above projection condition. The above computation also shows that constancy and covariant-constancy are equivalent for any spinor on the manifold $TN_4 \times T^{1,1}$ if the spinor obeys the projection condition.

\subsection*{Dimensional reduction of the Killing spinor equation}
On dimensionally reducing the theory along the $S^1$ fiber, we demand that $\partial_{\chi}\zeta=0$ --- automatically true for the chiral spinor which is a solution of the Killing spinor equation $D_{\chi} \zeta=0$. The spinors generating the supersymmetry transformations in the dimensionally reduced theory are therefore expected to be given by the remaining components the Killing spinor equation, viz.
$D_m \zeta=0$, with $m=r,\theta, \phi$. We will now express this equation in terms of the three-dimensional covariant derivative.

We will be interested in reducing the theory to flat 3D space, with the metric
\begin{equation}
\de s^2=\de r^2 + r^2 (\de\theta^2 + \sin^2{\theta} \de\phi^2)
\end{equation}
Choosing $e^1=\de r, e^2 = r \de\theta, e^3 = r\sin{\theta} \de \phi$, the spin connections for the above metric are
\begin{align}
&\Omega_{12}=-\de\theta,\\
&\Omega_{13}=-\sin{\theta}\,\de\phi,\\
&\Omega_{23}=-\cos{\theta}\,\de\phi.
\end{align}
Now, one can easily check that the equation for the Killing spinor $D_m \zeta=0$ on flat 3D space is identical to the dimensionally reduced Killing spinor equation on the Taub-NUT space for a spinor satisfying the constraint \ref{TN-KS}. The supersymmetry transformations for the 3D action are therefore generated by constant chiral spinors,
\begin{equation}
\partial_i \zeta=0,
\end{equation}
where $x^i$ are the Cartesian coordinates on $\mathbb{R}^3$.

\printbibliography

\end{document}